\documentclass[12pt]{article}
\usepackage[body={16cm, 23cm, centered}]{geometry}

\usepackage{
	amsmath,
	amssymb,
	amsfonts,
	amsthm,
	amscd,
	cite, 
	comment,
	indentfirst,
	epsfig,
	color,
	setspace,
	bbold,
	verbatim,
	MnSymbol,
	slashed
}

\usepackage[debug,pageanchor=false]{hyperref}
\hypersetup{colorlinks=true,linktocpage,breaklinks,
 	urlcolor=blue,
 	linkcolor=red,
 	citecolor=blue
 }
 
\numberwithin{equation}{section}

\def\a{\alpha} 
\def\b{\beta} 
 
\def\d{\delta}

\def\h{\eta}

\def\r{\rho}

\def\s{\sigma}

\def\P{\Pi}

\def\W{\Omega}

\def\bm{\bar{m}}

\def\fr{\frac}  \def\dt{\partial}

\def\ph{\phantom}

\def\mc{\mathcal}

\def\mF{\mathcal{F}}

\def\RR{\mathbb{R}}
\def\TT{\mathbb{T}}
\def\SS{\mathbb{S}}
\def\ZZ{\mathbb{Z}}

\def\bas{\mathrm{bas}\,}

\def\gg{\mathfrak{g}}
\def\tgg{\tilde{\mathfrak{g}}}

\newcommand{\Exp}[1]{\operatorname{e}^{#1}}
\newcommand{\diag}{\operatorname{diag}}
\newcommand{\abs}[1]{\lvert {#1} \rvert}
\newcommand{\rmd}{{\mathrm{d}}}
\newcommand{\nn}{\nonumber}
\newcommand{\Lie}{\pounds}

\newcommand{\cC}{\mathcal C}
\newcommand{\cF}{\mathcal F}
\newcommand{\cH}{\mathcal H}

\newcommand{\cM}{\mathcal M}

\newcommand{\sfa}{\mathsf{a}}
\newcommand{\sfb}{\mathsf{b}}
\newcommand{\sfc}{\mathsf{c}}

\begin{document}
\renewcommand{\contentsname}{}
\renewcommand{\refname}{\begin{center}References\end{center}}
\renewcommand{\abstractname}{\begin{center}\footnotesize{\bf Abstract}\end{center}} 

\renewcommand{\refname}{\begin{center}References\end{center}}

\begin{titlepage}
	
	\ph{1}	
	\vspace{3cm}

	\begin{center}
		\baselineskip=16pt
		{\Large \bf 
			Non-abelian U-duality at work
		}
		\vskip 1cm
			Edvard T. Musaev$^{a}$\footnote{\tt musaev.et@phystech.edu}, Yuho Sakatani$^{b}$\footnote{\tt yuho@koto.kpu-m.ac.jp }	
		\vskip .3cm
		\begin{small}
			{\it 
				$^a$Moscow Institute of Physics and Technology,
			    Institutskii per. 9, Dolgoprudny, 141700, Russia,\\
				$^b$Department of Physics, Kyoto Prefectural University of Medicine, \\
				1-5 Shimogamohangi-cho, Sakyo-ku, Kyoto, Japan
			}
		\end{small}
	\end{center}
		
	\vfill 
	\begin{center} 
		\textbf{Abstract}
	\end{center} 
	\begin{quote}
         Non-abelian U-duality originates from the construction of exceptional Drinfel'd algebra (EDA), which extends the constriction of the classical Drinfel'd double. This symmetry is a natural extension of Poisson--Lie T-duality and is believed to be a symmetry of Type II string/M-theory or their low-energy effective theories. In this paper, we consider non-abelian U-dualities of 11- or 10-dimensional backgrounds starting with E${}_{n(n)}$ EDA with $n\leq 6$ with vanishing trombone gauging. The latter guarantees that all dual backgrounds satisfy the standard supergravity equations of motion. In particular, when the duality includes a timelike T-duality, we obtain solutions of M$^*$-theory or Type II$^*$ background equations, as expected. Also starting with coboundary EDA's we provide examples of generalised Yang--Baxter deformations of M-theory and Type IIB backgrounds. The obtained results provide explicit examples when non-abelian U-duality works well as a solution generating transformation. 
	\end{quote} 
	\vfill
	\setcounter{footnote}{0}
\end{titlepage}
	
\clearpage
\setcounter{page}{2}

\tableofcontents

\section{Introduction}

One of the most important goals when thinking of string or M-theory is to find a set of vacua, good in a certain sense. As the example closest to phenomological applications one could think of searching for stable de Sitter vacua of string theory \cite{Danielsson:2018ztv}. In addition to the structure of a differential manifold (probably with torsion) these could include a variety of additional structures, such as orientifold planes, Dp-brane fluxes etc. \cite{Grana:2003ek,Blumenhagen:2006ci,Danckaert:2009hr}. Closer look at the spectrum of string or M-theory objects (equivalently maximal supergravities in 10 and 11 dimensions) shows that the conventional branes to not exhaust it. Concretely, one finds a huge set of branes whose tension behaves as $T \sim g_s{}^{-\a}$ with $\a$ being a natural number potentially not bound from above \cite{Obers:1998fb,deBoer:2010ud,Kleinschmidt:2011vu,Bergshoeff:2011ee, Bergshoeff:2012ex,deBoer:2012ma,Lombardo:2016swq} (for recent reviews see \cite{Musaev:2019zcr,Berman:2020tqn}). These branes, called exotic, generate supergravity backgrounds of very peculiar properties. In particular, some of these can be described in terms of T-folds, i.e. field configurations patched by T(U)-duality transformation, that mix metric and gauge degrees of freedom \cite{Hull:1994ys,Hull:2004in,Hull:2006qs}. As Dp-brane backgrounds can be characterised by fluxes of the corresponding $(p+2)$-form field strength, non-geometric backgrounds generated by exotic branes can be characterised by the so-called non-geometric fluxes. These are related to non-trivial monodromies of the background, originating from the patching \cite{Bergman:2007qq,Lust:2015yia}. 

Non-geometric fluxes can be naturally considered among the set of additional compactification data alongside fluxes of Ramond--Ramond (RR) fields and components of the torsion tensor, blowing up the variety of possible string vacua. Luckily enough, such defined set of string vacua is highly degenerate with respect to symmetries of the string, namely T- and U-duality. From the point of view of the string configurations of background fields, which could look very different and some of which could even violate equations of motion of the standard supergravity, are equivalent. Such relations between points in the space of vacua reduce the study to orbits of the corresponding symmetry groups. 

One starts with T-duality that is a symmetry of perturbative formulation of the string on a torus $\TT^d$. This can be shown to be represented by the group O$(d,d;\ZZ)$ acting on background fields. To show equivalence of the backgrounds related by a T-duality transformation one singles out a set of $d$ world-volume scalar fields corresponding to coordinates on the torus, gauges the abelian shift symmetry $U(1)^d$ of the background and then integrates out the Lagrange multiplier \cite{Buscher:1987qj,Duff:1989tf}. The latter has to be added to preserve the amount of d.o.f. when adding gauge fields. Integrating out all $d$ Lagrange multipliers introduces new $d$ scalar fields, usually referred to as dual coordinates. With minimal modifications the same procedure can be applied to group manifold backgrounds other than the abelian U(1)$^{d}$ and the corresponding transformation of background fields would be called non-abelian T-duality \cite{delaOssa:1992vci,Klimcik:1995ux,Klimcik:2002zj}. Both abelian and non-abelian T-duality transformations relate background field configurations, which are indistinguishable from the point of view of the string, and hence reduce the amount of inequivalent string vacua. 

Looking closely at abelian and non-abelian T-duality transformations one notices, that these have nice properties when written in terms of world-sheet N\"other currents \cite{Klimcik:1995ux,Klimcik:1995jn}. One finds, that the former starts with conserved currents that can be represented as closed 1-forms $dJ^I=0$ on the world-sheet and maps these into a set of currents with the same properties, i.e. $d \tilde{J}_I=0$\,. In contrast, the latter starts with conserved currents corresponding to 1-forms, that satisfy Maurer--Cartan equation $d J^I=\fr12 f_{JK}{}^I J^J \wedge J^K$, where $f_{IJ}{}^K$ are structure constants of the non-abelian isometry group of the background. Upon a non-abelian T-duality transformation these are mapped to currents, represented by closed 1-forms. Naturally, there exists a third option, where a non-closed current 1-form gets mapped to a non-closed 1-form. Upon certain algebraic restrictions relating structure constants in the original and dual Maurer--Cartan equations, the corresponding backgrounds are equivalent from the point of view of the string. Hence, this is a symmetry of the string and is usually referred to as Poisson--Lie (PL) T-duality or T-plurality \cite{Klimcik:1995ux,Klimcik:2002zj,VonUnge:2002xjf} (for recent reviews see \cite{Thompson:2019ipl,Demulder:2019bha}). More detailed description of the procedure is presented below in Section \ref{sec:PL}.

T-duality briefly described above is a perturbative symmetry, meaning that it holds order-by-order in string perturbation theory, though it is non-perturbative on the string world-sheet. However, the space of string vacua is mostly populated by 11-dimensional backgrounds corresponding to large string coupling, when degrees of freedom of the theory are better described by the membrane \cite{Hull:1994ys,Townsend:1995kk}. Low energy theory of the background fields in this case would be the 11-dimensional supergravity \cite{Cremmer:1978km,Bergshoeff:1987cm}. The perturbative T-duality symmetry along with the non-perturbative S-duality, relating Type IIB descriptions at $g_s$ and $1/g_s$, gets enhanced to a set of non-perturbative symmetries called U-dualities \cite{Hull:1994ys}. To some extent one may generalise the T-duality procedure based on integrating out the Lagrange multiplier to the case of membranes \cite{Duff:1990hn}, however, the full U-duality group of M-theory on an $n$-torus can be most transparently seen at the level of supergravity \cite{Cremmer:1997ct,Cremmer:1998px}. One observes, that U-duality group of 11-dimensional supergravity on a torus $\TT^n$ is given by split real forms of the E-series Lie algebras E${}_{n(n)}$, where the common notation reads
\begin{equation}
    \begin{aligned}
     & {\rm E}_{5(5)} = {\rm SO}(5,5), \\
     &{\rm E}_{4(4)} = {\rm SL}(5), \\
     & {\rm E}_{3(3)} = {\rm SL}(3)\times{\rm SL}(2),
    \end{aligned}
\end{equation}
and all groups are taken over real numbers. Fermions of the theory transform under the corresponding maximal compact subgroups.

Naturally one gets interested in an enhancement of the non-abelian T-duality and more generally PL T-duality symmetries to symmetries of 11-dimensional backgrounds. Rather expected, the direct generalisation of the analysis based on two-dimensional sigma-model to the case of the membrane is a tough path. Indeed, already for abelian U-duality symmetry, strictly speaking, one is not able to go beyond the SL(5) symmetry, since for that dynamics of the M5-brane has to be taken into account. Although some information on non-abelian generalisation of U-duality can be extracted from the membrane \cite{Sakatani:2020iad}, it proves fruitful to work in the low energy limit. In this work we focus at such generalisations of the abelian U-duality and work out examples of non-abelian U-duality pairs, which are not a simple lift of non-abelian T-duality pairs in 10 dimensions. For that we start with the machinery of PL U-duality or non-abelian U-duality developed in \cite{Sakatani:2019zrs,Malek:2019xrf,Sakatani:2020wah,Malek:2020hpo,Musaev:2020bwm}, which we review in necessary details in Section \ref{sec:PL}. Section \ref{sec:Uexmpls} presents the results classified by the corresponding abelian U-duality groups E${}_{n(n)}$. Finally, in Section \ref{sec:concl} we discuss the presented results and further developments.

\section{Non-abelian duality: User's manual}
\label{sec:PL}

For self-containment of the text and for further references we overview details of non-abelian T- and U-duality constructions. 
We follow the more general approach based on symmetry maps inside Drinfel'd double algebra and exceptional Drinfel'd algebra, often referred to as Poisson--Lie and Nambu--Lie dualities respectively. Let us first start with the PL T-duality procedure.

\subsection{Poisson--Lie T-duality}

PL T-duality can be understood twofold: algebraically and geometrically, where the former picture is based on equivalence of different parametrisations of the same Drinfel'd double algebra in terms of Manin triples, while the latter additionally provides geometric realisation for every parametrisation hence establishing a duality correspondence between string backgrounds. In a moment we will discuss both these descriptions.

\subsubsection{Algebra}

From the algebraic perspective PL T-duality symmetry is based on the classical Drinfel'd double defined as an even dimensional Lie algebra, that admits decomposition in terms of Manin triple $(\mc{D},\gg,\tgg)$, where the algebra $\mc{D}$ is endowed by a symmetric bilinear form $\h$ and $\gg$ and $\tgg$ are isotropic subalgebras such that $\mc{D}=\gg \oplus \tgg$ as a linear vector space. Specifically, one defines generators $\{T_a\}=\bas\, \gg$, $\{T^a\}=\bas\,\tgg$, action of the symmetric bilinear form $\h(T_a,T^b) = \d_a{}^b$ and the following commutation relations
\begin{equation}
\label{eq:DD}
    \begin{aligned}{}
        [T_a,\,T_b] & = f_{ab}{}^c\,T_c\,, \quad
        [T^a,\,T^b] = f_c{}^{ab}\,T^c\,, \\
         [T_a,\,T^b] &= f_a{}^{bc}\, T_c - f_{ac}{}^b\, T^c\,.
    \end{aligned}
\end{equation}
Here $a=1,\dotsc,d$ and $d$ is some integer number that will further be equal to the dimension of the (group manifold) background.

Generators of the algebra can be denoted collectively as $T_A=(T_a,\,T^a)$ with $A=1,\dotsc,2d$\,, that allows to write commutation relations as $[T_A,\,T_B]=\cF_{AB}{}^C\,T_C$ and the symmetric bilinear form as
\begin{equation}
\label{eq:eta}
 \h( T_A,\,T_B) = \eta_{AB}\,,\qquad
 \eta_{AB}=\begin{pmatrix} 0 & \delta_a^b \\ \delta^a_b & 0 \end{pmatrix}.
\end{equation}
Structure constants $\cF_{AB}{}^C$ satisfy
\begin{equation}
    \h( [T_A,\,T_B],\,T_C) +  \h( T_B,\,[T_A,\,T_C]) = 0\,,
 \end{equation}
which in components means antisymmetry in the last two indices of $\cF_{ABC}\equiv \cF_{AB}{}^D\,\eta_{DC}$\,.

Crucial point here is that a given Drinfel'd algebra can in general be realised by multiple choice of Manin triples, related by a transformation that would be called PL T-duality (plurality). For explicit examples one can refer to the classification of all six-dimensional real Drinfel'd doubles in terms of Manin triples \cite{Snobl:2002kq}. It is worth mentioning, that some of these have more than two realisations, while some have only one. 

\subsubsection{Geometry}
\label{sec:plgeom}

Let us now turn to geometric realisation of Drinfel'd double, that relates the above construction to symmetries of string backgrounds. For that we perform the following steps.

\textbf{1. }For a given Drinfel'd double pick out a subalgebra $\mathfrak{g}$ that is maximally isotropic with respect to the bilinear form $\h$\,. For a given realisation the canonical choice of the subalgebra is spanned by $\{T_a\}$, which indeed forms a subalgebra $[T_a,\,T_b] = f_{ab}{}^c\,T_c$, and satisfies $\h(T_a,\,T_b)=0$\,. 

\textbf{2.}
Construct a group element of $G=\exp\mathfrak{g}$ (e.g. as $g=\Exp{x^a\,T_a}$) and define the right-invariant 1-forms $r^a=r_m{}^a dx^m$ and vector fields $e_a=e_a{}^m \dt_m$ as\footnote{The index $m=1,\dotsc,d$ plays the same role as the index $a$. When it is interpreted as the curved index we denote the index as $m$ while we use $a$ for the ``flat'' index.}
\begin{align}
 r\equiv r_m{}^a\,\rmd x^m\,T_a = \rmd g\,g^{-1}\,,\qquad
 r^a(e_b)=\d_b{}^a\,.
\end{align}

\textbf{3.} 
Compute the adjoint action of $g^{-1}$ on generators $T_A$,
\begin{align}
 g^{-1}\,T_A\,g \equiv M_A{}^B(x)\,T_B\,.
\end{align}
It turns out that the matrix $M_A{}^B(x)$ is an element of $\text{O}(d,d)$ and, in particular, can be parameterized as \cite{Klimcik:1995ux}
\begin{align}
 M_A{}^B(x) = \begin{pmatrix} \delta_a^c & 0 \\ -\pi^{ac}(x) & \delta^a_c \end{pmatrix} \begin{pmatrix} a_c{}^b(x) & 0 \\ 0 & (a^{-1}(x))_b{}^c \end{pmatrix},
\end{align}
where $\pi^{ab}=-\pi^{ba}$\,. 

\textbf{4.} 
Define generalised frame fields as
\begin{align}
 E_A{}^I(x) \equiv \begin{pmatrix} \delta_a^b & 0 \\ -\pi^{ab} & \delta^a_b \end{pmatrix}
 \begin{pmatrix} e_b^m & 0 \\ 0 & r_m^b \end{pmatrix}.
\end{align}
The important property of such defined generalised frame fields is that these satisfy the algebra \cite{Hassler:2017yza,Demulder:2018lmj}
\begin{align}
 \hat{\pounds}_{E_A}E_B{}^I = - \cF_{AB}{}^C\,E_C{}^I\,,
\end{align}
where $\cF_{AB}{}^C$ is the structure constants of the initial Drinfel'd double and $\hat{\pounds}_{V}$ denotes the generalised Lie derivative of double field theory. 
For two generalised vectors $V^I=(v^m,\,\tilde{v}_m)$ and $W^I=(w^m,\,\tilde{w}_m)$, the latter is defined as
\begin{align}
 \hat{\pounds}_{V}W^I = \begin{pmatrix} \pounds_v w^m \\
 \pounds_v \tilde{w}_m - \iota_w \rmd \tilde{v}_m \end{pmatrix} .
\end{align}

\textbf{5.} 
Introduce a constant matrix $\hat{\cH}_{AB}\in \text{O}(d,d)$,\footnote{In general, this constant matrix can depend on other coordinates, called the spectator fields (see \cite{Klimcik:1995ux} and also \cite{Sakatani:2019jgu}).} and define the generalised metric as
\begin{align}
 \cH_{IJ}(x) \equiv E_I{}^A(x)\,E_J{}^B(x)\,\hat{\cH}_{AB}\,,
\end{align}
where $E_I{}^A(x)$ denote the inverse of $E_A{}^I$\,. 
Using the standard parametrisation
\begin{align}
 \cH_{IJ}(x) = \begin{pmatrix}
 (g+B\,g^{-1}\,B)_{mn} & (B\,g^{-1})_{m}{}^n \\ -(g^{-1}\,B)^m{}_n & g^{mn} \end{pmatrix},
\label{eq:cH-param}
\end{align}
of the generalised metric one identifies the supergravity fields $(g_{mn},\,B_{mn})$\,.
The dilaton and the RR fields also can be constructed by using the objects $\{e_a^i,\,a_a{}^b,\,\pi^{ab}\}$ although we do not consider these here (see \cite{VonUnge:2002xjf,Hassler:2017yza,Demulder:2018lmj,Sakatani:2019jgu} for more detail).

Following the above steps one obtains a supergravity background from a given Drinfel'd algebra. If the background has vanishing dilaton and the RR-fluxes, Type II supergravity equations of motion reduce to \cite{Demulder:2018lmj} (see also \cite{Geissbuhler:2013uka})
\begin{align}
\begin{split}
\cF_{ABC}\,\cF_{DEF} \,\bigl(3\,\hat{\cH}^{AD}\,\eta^{BE}\,\eta^{CF}- \hat{\cH}^{AD}\,\hat{\cH}^{BE}\,\hat{\cH}^{CF}\bigr) &= 0 \,,
\\
(\eta^{CE}\,\eta^{DF} - \hat{\cH}^{CE}\,\hat{\cH}^{DF} \bigr)\,\hat{\cH}^{G[A}\,\cF_{CD}{}^{B]}\,\cF_{EFG} &= 0 \,, 
\end{split}
\label{eq:DFT-EOM}
\end{align}
where $\eta^{AB}$ is the inverse of $\eta_{AB}$\,. 
Then, if $\hat{\cH}_{AB}$ is chosen such that these equations are satisfied one obtains a supergravity solution from a given Drinfel'd algebra. In principle, a different choice of the isotropic subalgebra inside a given Drinfel'd algebra will render a different background, however, since the constant metric $\hat{\cH}_{AB}$ and structure constants $\cF_{AB}{}^C$ are simply transformed by an O$(d,d)$ rotation, equations of motion will be equivalently satisfied. Let us describe the procedure of dualisation in more details.

\subsubsection{Duality}

Algorithm for producing a PL T-dual background from a given geometric realisation of a Drinfel'd double goes along the following lines. 

\textbf{1.} From the same Drinfel'd double pick out a different maximally isotropic subalgebra $\mathfrak{g}'$\, 
and identify an $\text{O}(d,d)$ transformation
\begin{align}
 T_A\to T'_A= C_A{}^B\,T_B\,,
\label{eq:PL-Odd}
\end{align}
such that the subalgebra $\mathfrak{g}'$ is generated by $\{T'_a\}$\,. 
Under this rotation, the structure constants are rotated as
\begin{align}
 \cF_{AB}{}^C \to \cF'_{AB}{}^C = C_A{}^D\,C_B{}^E\,(C^{-1})_F{}^C\,\cF_{DE}{}^F\,. 
\end{align}
Here, we assume that components $\cF'^{abc}$ are absent. 
Then the new algebra $\cF'_{AB}{}^C$ can be regarded as the Drinfel'd double \eqref{eq:DD} with new structure constants $f'_{ab}{}^c$ and $f'_c{}^{ab}$\,. 

\textbf{2.} The constant metric $\hat{\cH}_{AB}$ also rotates under the transformation
\begin{align}
 \hat{\cH}_{AB} \to \hat{\cH}'_{AB} = C_A{}^C\,C_B{}^D\,\hat{\cH}_{CD}\,.
\end{align}
Since the equations of motion \eqref{eq:DFT-EOM} are covariant under $\text{O}(d,d)$, the new choice $\hat{\cH}'_{AB}$ and $\cF'_{AB}{}^C$ also satisfy \eqref{eq:DFT-EOM} if the original ones satisfy these. 

\textbf{3.} Using the redefined generators, construct the generalised frame fields $E'_A{}^I$ and the generalised metric $\cH'_{IJ}=E'_I{}^A\,E'_J{}^B\,\hat{\cH}'_{AB}$ as before. 
Using the parametrisation \eqref{eq:cH-param} one identifies the dual supergravity fields $(g'_{mn},\,B'_{mn})$\,.

Following these steps, we can map a supergravity solution to its dual. 
The point is that the Drinfel'd double allows for a different choice of the maximally isotropic subalgebra each corresponding to a supergravity (double field theory) solution. 
For the particular choice,
\begin{align}
 C_A{}^B = \begin{pmatrix} 0 & \delta_a^b \\ \delta^a_b & 0 \end{pmatrix},
\label{eq:C-PL-T}
\end{align}
which simply exchanges $\gg$ with $\tgg$ this duality is called the PL T-duality \cite{Klimcik:1995ux}.
Additionally, if $f_{ab}{}^c=0$ or $f_c{}^{ab}=0$ the transformation is referred to as the non-abelian T-duality. 
In general, one finds multiple choices of $C_A{}^B$ and then the original solution can produce several backgrounds. 
In that case, the duality is called the PL T-plurality \cite{VonUnge:2002xjf}. 
For example, in the case of $d=3$, the inequivalent choices of $C_A{}^B$ have been classified in \cite{Snobl:2002kq} for each Drinfel'd double. 

It is important to mention that the bi-vector field
\begin{align}
 \pi \equiv \frac{1}{2!}\,\pi^{mn}\,\partial_m\wedge \partial_n \equiv \frac{1}{2!}\,\pi^{ab}\,e_a\wedge e_b\,,
\end{align}
satisfies
\begin{equation}
\begin{aligned}
 &\pi^{qm}\,\partial_q\pi^{np} + \pi^{qn}\,\partial_q\pi^{pm} + \pi^{qp}\,\partial_q\pi^{mn} = 0\,,\\
 &\pounds_{v_a}\pi^{mn} = f_a{}^{bc}\,v_b^m\,v_c^n\,,
\end{aligned}    
\end{equation}
where $v_a^m=(a^{-1})_a{}^b\,e_b^m$ is the left-invariant vector field with the property $\pounds_{v_a}v_b=f_{ab}{}^c\,v_c^m$\,. 
This allows to refer to the bi-vector field $\pi$ as to a Poisson--Lie structure. In the case of non-abelian U-duality which we will discuss immediately, the Poisson--Lie structure is extended to a Nambu--Lie structure \cite{Vaisman2000}. 
Accordingly the transformation may be called the Nambu--Lie U-duality. 

\subsection{Nambu--Lie \emph{U}-duality}

The procedure of PL T--duality symmetry of 10-dimensional supergravity backgrounds (and more generally, of backgrounds of double field theory) can be generalised to transformations of 11-dimensional backgrounds. This uplifts the abelian U-duality symmetry of maximal supergravity to a set of transformations that can naturally be called Nambu--Lie U-dualities. One however faces certain new complexities inherent in U-duality, which we explain in detail immediately.

\subsubsection{Algebra}

Instead of the Drinfel'd double Lie algebra the procedure of Nambu--Lie U-duality is based on the construction of exceptional Drinfel'd algebra (EDA), which is a Leibniz algebra whose structure is highly restricted by a set of constraints. As for Drinfel'd doubles it is convenient to describe EDA's in terms of a generalisation of Manin triples, i.e. a subalgebra $\gg$, its linear complement $\tgg$ inside the EDA and a generalisation of the map $\h: D \times D \to \RR$, that is
\begin{equation}
    \label{eq:sc}
    \gg \otimes \gg\, \big|_{\mc{R}_2} = 0\,,
\end{equation}
where $\mc{R}_2$ is a certain representation of the corresponding abelian U-duality group, which for the dimensions $n=\dim \gg$ read
\begin{equation}
    \begin{alignedat}{3}
     & n=4\,, &\quad& \mc{R}_2 = \mathbf{\overline{5}} &\quad&\text{of \, SL(5)}\,, \\
     & n=5\,, && \mc{R}_2 = \mathbf{10} &&\text{of \, SO(5,5)}\,, \\
     & n=6\,, && \mc{R}_2 = \mathbf{\overline{27}} &&\text{of \, E}_{6(6)}\,. 
    \end{alignedat}
\end{equation}
This linear constraint can be thought of as a condition defining what would be called the isotropic subalgebra. Closer analysis of the condition shows, that unlike the Drinfel'd double, the EDA contains two different types of maximally isotropic subalgebras\footnote{This has the same origin as the two types of solutions to the section condition of exceptional field theory}. The first one has at most dimension $n$ and corresponds to M-theory backgrounds, while the second has at most dimension $(n-1)$ corresponds to Type IIB backgrounds. 

Denoting generators of such defined isotropic subalgebra $\{T_a\}=\bas \gg$, one denotes the rest $\{T^{a_1a_2}, T^{a_1,\dots a_5}\}$ and defines the corresponding EDA as a set of multiplication rules
\begin{equation}
    T_A\circ T_B=\mF_{AB}{}^C T_C,
\end{equation}
where we collectively denote the generators as $T_A=\{T_a,T^{a_1a_2}, T^{a_1,\dots a_5}\}$ and $\mF_{AB}{}^C$ denote the structure constants. The product $\circ$ satisfies the Leibniz identity 
\begin{align}
 a \circ (b\circ c) = (a\circ b)\circ c + b\circ (a\circ b) \,,
\end{align}
rendering the EDA a Leibniz algebra. In general, this product is not skew-symmetric and the EDA is not a Lie algebra. Explicit form of the multiplication table for $n\leq 6$ is presented in  \eqref{eq:eda6}.

For $n<6$ the EDA can be obtained simply by restricting the indices to run over $a=1,\dotsc,n$\,. For example, when $n=4$, the generator $T^{a_1\cdots a_5}$ disappears and we obtain the E${}_{4(4)}$ EDA \cite{Sakatani:2019zrs,Malek:2019xrf}
\begin{align}
\begin{split}
 T_a \circ T_b &= f_{ab}{}^c\,T_c \,,
\\
 T_a \circ T^{b_1b_2} &= f_a{}^{b_1b_2c}\,T_c + 2\,f_{ac}{}^{[b_1}\,T^{b_2]c}
 +3\,Z_a\,T^{b_1b_2}\,,
\\
 T^{a_1a_2} \circ T_b &= -f_b{}^{a_1a_2c}\,T_c + 3\,f_{[c_1c_2}{}^{[a_1}\,\delta^{a_2]}_{b]}\,T^{c_1c_2}\\
 &\quad
 -9\,Z_c\,\delta_b^{[c}\,T^{a_1a_2]}\,,
\\
 T^{a_1a_2} \circ T^{b_1b_2} &= -2\, f_c{}^{a_1a_2[b_1}\, T^{b_2]c} \,,
\end{split}
\end{align}
which is 10 dimensional. 
Taking $n=7,8$ additional generators and structure constants appear and the structure gets more complicated (see \cite{Sakatani:2020wah} for the detail).

In these notations for $n\leq 6$ the condition \eqref{eq:sc} can be written by defining 
\begin{align}
 \langle T_A,\,T_B\rangle^{\cC} = \eta_{AB}{}^{\cC}\,,
\end{align}
where $\cC$ is the index labelling the irrep $\mc{R}_2$ of the E${}_{n(n)}$ group. The tensor $\eta_{AB}{}^{\cC}$ generalises the metric \eqref{eq:eta} of the Drinfel'd double and is simply the invariant tensor connecting the representations $\mc{R}_1\times_{\text{sym}} \mc{R}_1$ and $R_2$\,. One finds that the subalgebra generated by $\{T_a\}$ is maximally isotropic with respect to this metric
\begin{align}
 \langle T_a,\,T_b\rangle^{\cC} = 0\,, 
\end{align}
that is precisely the condition \eqref{eq:sc} written in terms of generators.

Choosing an $(n-1)$-dimensional maximally isotropic algebra which is not subalgebra of the n-dimensional algebra spanned by $\{T_a\}$, one recovers the so-called Type IIB exceptional Drinfeld algebra. For $n\leq 6$ its multiplication table is explicitly presented in \eqref{eq:edaB}. The name is for the Type IIB theory symmetry of whose backgrounds it encodes. In fact, inspecting explicit examples one finds that the M-theory EDA can be mapped to the Type IIB EDA via a certain change of generators (see \cite{Sakatani:2017nfr} for such a map and also \cite{Sakatani:2019zrs,Blair:2020ndg} for some examples). 

\subsubsection{Geometry}

The construction of Section \ref{sec:plgeom} admits straightforward generalisation to the case of exceptional Drinfel'd algebras. Hence, for a given EDA one constructs a supergravity background following the procedure below.

\textbf{1.} From the EDA pick out a subalgebra $\mathfrak{g}$ that is maximally isotropic given the tensor $\langle\cdot,\,\cdot\rangle^{\cC}$\,. 
The subalgebra $\mathfrak{g}$ turns out to be a Lie algebra, and one denotes the generators as $\{T_a\}$\,, where $a=1,\dotsc,n$ in M-theory while $a=1,\dotsc,n-1$ in Type IIB theory. 

\textbf{2.} Construct a group element of $G=\exp\mathfrak{g}$, for example, as $g=\Exp{x^a\,T_a}$ and define the right-invariant 1-forms $r_i^a$ and right-invariant vector fields $e_a^i$ as
\begin{align}
 r\equiv r_i^a\,\rmd x^i\,T_a = \rmd g\,g^{-1}\,,\qquad
 e_a^i \equiv (r_i^a)^{-1}\,.
\end{align}

\textbf{3.} 
Compute the ``adjoint action'' of $g^{-1}\equiv \Exp{h}$ on $T_A$,
\begin{equation}
\begin{aligned}
 g^{-1}\triangleright T_A &\equiv T_A + h\circ T_A + \tfrac{1}{2!}\,h\circ (h\circ T_A) \\
 &\quad + \tfrac{1}{3!}\,h\circ (h\circ (h\circ T_A)) +\cdots
\\
 &\equiv M_A{}^B(x)\,T_B\,.
\end{aligned}
\end{equation}
It turns out that the matrix $M_A{}^B(x)$ is an element of E${}_{n(n)}$ and, in particular, can be parameterized as
\begin{align}
 M_A{}^B(x) = \mathbf{\Pi}_A{}^C(x)\,A_C{}^B(x)\,.
\end{align}

For the M-theory solution of the linear constraint the matrices $\mathbf{\Pi}_A{}^C$ and $A_A{}^B$ can be written explicitly in components as \cite{Sakatani:2019zrs,Malek:2019xrf,Malek:2020hpo}
\begin{align}
 \mathbf{\Pi}_A{}^B &= \begin{pmatrix}
 \delta_a^b & 0 & 0 \\ -\frac{1}{\sqrt{2!}}\pi^{ba_1a_2} & \delta_{b_1 b_2}^{a_1 a_2}& 0 \\
 \P^{ba_1\cdots a_5} & \frac{20}{\sqrt{2!\,5!}}\,\delta_{b_1b_2}^{[a_1a_2}\,\pi^{a_3a_4a_5]} & \delta_{b_1\ldots b_5}^{a_1\dots a_5} \end{pmatrix},
\\
 A_A{}^B &= \begin{pmatrix}
 a_a{}^b & 0 & 0 \\ 
 0 & (a^{-1})_{b_1b_2}^{a_1a_2} & 0 \\
 0 & 0 & (a^{-1})_{b_1\dots b_5}^{a_1\dots a_5}\end{pmatrix} ,
\end{align}
where we define 
\begin{equation}
    \begin{aligned}
     \P^{ba_1\cdots a_5}&=-\frac{\pi^{ba_1\cdots a_5} + 5\,\pi^{b[a_1a_2}\,\pi^{a_3a_4a_5]}}{\sqrt{5!}},\\
     \delta_{b_1\ldots b_p}^{a_1\dots a_p}&=\delta_{b_1}^{[a_1}\cdots \d_{b_p}^{a_p},\\
     (a^{-1})_{b_1\dots b_q}^{a_1 \dots a_q} &=(a^{-1})_{[b_1}{}^{a_1}\dots (a^{-1})_{b_q]}{}^{a_q}.
    \end{aligned}
\end{equation}

For the Type IIB solution (with $n\leq 5$), the parameterizations read \cite{Sakatani:2020wah}
\begin{align}
 \mathbf{\Pi}_A{}^B &= \begin{pmatrix}
 \delta_{\sfa}^{\sfb} & 0 & 0 \\ - \pi_\alpha^{\sfb\sfa} & \delta_\alpha^\beta\,\delta^{\sfa}_{\sfb} & 0 \\
  \P^{\sfa_1\sfa_2\sfa_3} & \frac{3\,\epsilon^{\beta\gamma}\,\delta^{[\sfa_1}_{\sfb}\,\pi_\gamma^{\sfa_2\sfa_3]}}{\sqrt{3!}} & \delta_{\sfb_1\sfb_2\sfb_3}^{\sfa_1\sfa_2\sfa_3}\end{pmatrix},
\\
 A_A{}^B &= \begin{pmatrix}
 a_{\sfa}{}^{\sfb} & 0 & 0 \\ 0 & \lambda_\alpha{}^{\beta}\,(a^{-1})_{\sfb}{}^{\sfa} & 0 \\
 0 & 0 & (a^{-1})_{\sfb_1 \dots \sfb_3}^{\sfa_1 \dots \sfa_3} \end{pmatrix} ,
\end{align}
where $\epsilon^{\mathbf{1}\mathbf{2}}=\epsilon_{\mathbf{1}\mathbf{2}}=1$\, and $\P^{\sfa_1\sfa_2\sfa_3}=-\frac{1}{\sqrt{3!}}\pi^{\sfa_1\sfa_2\sfa_3}+\frac{3}{2}\,\epsilon^{\gamma\delta}\,\pi_\gamma^{\sfb[\sfa_1}\,\pi_\delta^{\sfa_2\sfa_3]}$.  The indices $\sfa,\sfb$ run over $\sfa,\sfb=1,\dotsc,n-1$ and $\alpha,\beta =\mathbf{1},\mathbf{2}$ label doublets of the $\text{SL}(2)$ $S$-duality.

\textbf{4.}
Given the matrix $\mathbf{\Pi}_A{}^C$ define the generalised frame fields as
\begin{align}
 E_A{}^I(x) \equiv \mathbf{\Pi}_A{}^B(x)\,\mathbb{E}_B{}^I(x)\,,
\end{align}
where $\mathbb{E}_A{}^I$ is defined as
\begin{equation}
\begin{aligned}
 \mathbb{E}_A{}^I &= \begin{pmatrix}
 e_a^i & 0 & 0 \\ 0 & \Exp{-3\Delta}r_{[i_1}^{a_1}\,r_{i_2]}^{a_2} & 0 \\
 0 & 0 & \Exp{-6\Delta}r_{[i_1}{}^{a_1}\cdots r_{i_5]}{}^{a_5} \end{pmatrix}, \\
 &\mbox{for M-theory, i.e. } n\leq 6\,, \\ \\
 \mathbb{E}_A{}^I &=\begin{pmatrix}
 e_{\sfa}^m & 0 & 0 \\ 0 & \Exp{-2\Delta}\lambda_\alpha{}^{\beta}\,r_{m}^{\sfa} & 0 \\
 0 & 0 & \Exp{-4\Delta}r_{[m_1}{}^{\sfa_1}\,r_{m_2}{}^{\sfa_2}\,r_{m_3]}{}^{\sfa_3} \end{pmatrix} , \\
 &\mbox{for Type IIB, i.e. } n\leq 5\,.
\end{aligned}
\end{equation}
As it has been shown explicitly in \cite{Sakatani:2019zrs,Malek:2019xrf,Malek:2020hpo,Sakatani:2020wah}, such defined set of generalised frame fields $E_A{}^I$ satisfies the algebra
\begin{align}
 \hat{\pounds}_{E_A}E_B{}^I = - \mF_{AB}{}^C\,E_C{}^I\,,
\end{align}
where $\mF_{AB}{}^C$ denote structure constants of the EDA and $\hat{\pounds}_{V}$ denotes the generalised Lie derivative of exceptional field theory. 

\textbf{5.}
Introduce a constant matrix $\hat{\cM}_{AB}\in E_{n(n)}\times \mathbb{R}^+$ and define the generalised metric as
\begin{align}
 \cM_{IJ}(x) \equiv E_I{}^A(x)\,E_J{}^B(x)\,\hat{\cM}_{AB}\,,
\end{align}
where $E_I{}^A(x)$ denote the inverse matrix of $E_A{}^I$\,. 
As is well known in exceptional field theory (see for example \cite{Berman:2011jh}), the generalised metric can be decomposed as
\begin{align}
 \cM_{IJ}(x) = L_I{}^K(x)\,L_J{}^L(x)\,\hat{\cM}_{KL}(x)\,.
\label{eq:gen-met}
\end{align}
Where the matrices $L_I{}^J$ and $\hat{\cM}_{IJ}$ are parametrized in terms of background fields. For the M-theory solutions of the linear constraint the parameterization reads
\begin{equation}
\begin{aligned}
 \hat{\cM}_{IJ} &= (\det g)^{\frac{1}{9-n}} \begin{pmatrix}
 g_{ij} & 0 & 0 \\
 0 & g^{i_1i_2,j_1j_2} & 0 \\
 0 & 0 & g^{i_1\cdots i_5,j_1\cdots j_5} 
\end{pmatrix},
\\
 L_I{}^J &=
 \begin{pmatrix}
 0 & \frac{A_{ij_1j_2}}{\sqrt{2!}} & \frac{A_{ij_1\cdots j_5}-5\,A_{i[j_1j_2}\,A_{j_3j_4j_5]}}{\sqrt{5!}} \\
 0 & 0 & -\frac{20\,\delta^{i_1i_2}_{[j_1j_2}\,A_{j_3j_4j_5]}}{\sqrt{2!\,5!}}\\
 0 & 0 & 0 
 \end{pmatrix},
\end{aligned}
\end{equation}
where $g^{i_1\cdots i_p,j_1\cdots j_p}\equiv g^{i_1k_1}\cdots g^{i_pk_p}\,\delta_{[k_1}^{j_1}\cdots \delta_{k_p]}^{j_p}$\,. 
For the Type IIB solution of the linear constraint these are given by
\begin{equation}
 \hat{\cM}_{IJ} = (\det \mathsf{g})^{\frac{1}{9-n}} \begin{pmatrix}
 \mathsf{g}_{mn} & 0 & 0 \\
 0 & m_{\alpha\beta}\,\mathsf{g}^{mn} & 0 \\
 0 & 0 & \mathsf{g}^{m_1m_2m_3,n_1n_2n_3} 
 \end{pmatrix},
\end{equation}
\begin{equation}
 L_I{}^J =
 \begin{pmatrix}
 0 & B^\beta_{mn} & \frac{D_{m_1m_2m_3}-\frac{3}{2}\,\epsilon_{\gamma\delta}\,B^\gamma_{m[n_1}\,B^{\delta}_{n_2n_3]}}{\sqrt{3!}} \\
 0 & 0 & -\frac{3\,\epsilon_{\alpha\gamma}\,\delta_{[n_1}^m B^{\gamma}_{n_2n_3]}}{\sqrt{3!}} \\
 0 & 0 & 0 
\end{pmatrix}.
\end{equation}
Here, $\mathsf{g}_{mn}$ is the Einstein-frame metric and $m_{\alpha\beta}$ an element of the coset SL(2)/SO(2) and is usually taken to be
\begin{align}
 m_{\alpha\beta} = \begin{pmatrix} 1 & -C_0 \\ 0 & 1 \end{pmatrix}
 \begin{pmatrix} \Exp{-\Phi} & 0 \\ 0 & \Exp{\Phi} \end{pmatrix}
 \begin{pmatrix} 1 & 0 \\ -C_0 & 1 \end{pmatrix}.
\end{align}
The $S$-duality doublet $B^\alpha_{mn}$ and the singlet $D_{m_1\cdots m_4}$ are defined as
\begin{equation}
\begin{aligned}
 B^\alpha_{mn} &= \begin{pmatrix} B^{\mathbf{1}}_{mn} \\ B^{\mathbf{2}}_{mn} \end{pmatrix} = \begin{pmatrix} B_{mn} \\ -C_{mn} \end{pmatrix},\\
 D_{m_1\cdots m_4} &= C_{m_1\cdots m_4} + 3\,C_{[m_1m_2}\,B_{m_3m_4]}\,.
\end{aligned}
\end{equation}
This parameterization allows to identify supergravity fields from the constructed generalised metric $\cM_{IJ}$\,. 

\textbf{6.} For the case of Nambu--Lie U-duality one must perform one more step. In addition to the $n$- or $(n-1)$-dimensional (say, internal) components of the supergravity fields we have considered so far, one has $d\equiv 11-n$ external components as well as mixed components. 
Unlike T-duality, these additional components are not singlets under U-duality, and one has to consider their transformation rules in order to generate a supergravity solution. 
 
In this paper, for simplicity, we consider the case where only the external metric $g_{\mu\nu}$ ($\mu,\nu=1,\dotsc,d$) is non-trivial, while the other fields vanish. The rescaled metric
\begin{align}
\begin{split}
 \text{\underline{M-theory}}:&\qquad \mathfrak{g}_{\mu\nu} \equiv \abs{\det g_{ij}}^{\frac{1}{9-n}}\,g_{\mu\nu}\,,
\\
 \text{\underline{Type IIB}}:&\qquad \mathfrak{g}_{\mu\nu} \equiv \abs{\det \mathsf{g}_{mn}}^{\frac{1}{9-n}}\,\mathsf{g}_{\mu\nu}\,,
\end{split}
\end{align}
also known as `the external metric of exceptional field theory' is invariant under global U-duality transformations (see \cite{Hohm:2013vpa} for details and \cite{Bakhmatov:2020kul,Gubarev:2020ydf} for the corresponding transformations under tri- and six-vector deformations). 

In all of the examples below the initial background $g_{ij}$ is considered to be a Ricci flat space with constant determinant. 
We can thus choose $\mathfrak{g}_{\mu\nu}$ to be a constant metric.

\subsubsection{Duality}

The procedure of dualisation follows the same steps as that for the PL T-duality.

\textbf{1.} From a given EDA consider a redefinition of generators $T_A\to T'_A= C_A{}^B\,T_B$ such that the new generators $T'_A$ enjoy the M-theory or Type IIB EDA. 
Under this redefinition, the structure constants become
\begin{align}
 \mF_{AB}{}^C \to \mF'_{AB}{}^C = C_A{}^D\,C_B{}^E\,(C^{-1})_F{}^C\,\mF_{DE}{}^F\,. 
\end{align}

\textbf{2.} Rotate the constant metric $\hat{\cM}_{AB}$ accordingly
\begin{align}
 \hat{\cM}_{AB} \to \hat{\cM}'_{AB} = C_A{}^C\,C_B{}^D\,\hat{\cM}_{CD}\,.
\end{align}

\textbf{3.} Using the redefined generators construct the generalised frame fields $E'_A{}^I$ and the generalised metric $\cM'_{IJ}=E'_I{}^A\,E'_J{}^B\,\hat{\cM}'_{AB}$\,. 
Identify dual supergravity fields. In particular the external metric becomes
\begin{align}
\begin{split}
 \text{\underline{M-theory}}:&\qquad g'_{\mu\nu} \equiv \abs{\det g'_{ij}}^{-\frac{1}{9-n}}\,\mathfrak{g}_{\mu\nu}\,,
\\
 \text{\underline{Type IIB}}:&\qquad \mathsf{g}'_{\mu\nu} \equiv \abs{\det \mathsf{g}'_{mn}}^{-\frac{1}{9-n}}\,\mathfrak{g}_{\mu\nu}\,,
\end{split}
\end{align}

In contrast to the case of 10-dimensional backgrounds, the equations of motion of exceptional field theory have not been expressed in a similar form as \eqref{eq:DFT-EOM}.
Therefore, in principle it has not been proven that the dual background is a solution of supergravity. However, inspecting the Lagrangian of exceptional field theory one could expect that the scalar sector together with the external metric work precisely as in the case of flux formulation of double field theory. Moreover, at least for U-duality groups smaller than E${}_{7(7)}$, where the additional self-duality constraint has to be imposed, one does not expect any peculiar behaviour. Keeping explicit proof of the solution generating nature of the above procedure to a future work, we focus here at explicit examples. We observe, that all of the dual background obtained via the Nambu--Lie dualisation are solutions to supergravity equations of motion, which suggests that non-abelian U-duality is a symmetry of supergravity equations of motion.\footnote{In this paper, we consider only examples where EDA is unimodular $X_{AB}{}^B=0$\,. In the non-unimodular case, dual background may be a solution of certain gauged supergravity (with the trombone gauging turned on \cite{LeDiffon:2008sh}).}

Similar to the Poisson--Lie structure one is able to define a tri-vector field
\begin{align}
 \pi \equiv \frac{1}{3!}\,\pi^{ijk}\,\partial_i\wedge \partial_j\wedge \partial_k \equiv \frac{1}{3!}\,\pi^{abc}\,e_a\wedge e_b\wedge e_c\,.
\end{align}
This satisfies the following properties
\begin{equation}
\begin{aligned}
 0&=\pi^{a_1a_2 d}\, \nabla_d \pi^{b_1b_2c} - 3\, \pi^{d[b_1b_2}\,\nabla_d \pi^{c]a_1a_2}\\
 &\quad 
  - f_{d_1d_2}{}^{[a_1}\, \pi^{a_2] b_1b_2 c d_1d_2}  \,,
\\
 \Lie_{v_a}\pi^{i_1i_2i_3} &= \Exp{-3 \Delta} f_{a}{}^{b_1b_2b_3}\,v_{b_1}^{i_1}\,v_{b_2}^{i_2}\,v_{b_3}^{i_3}\,,
\end{aligned}
\end{equation}
where $\nabla_b\pi^{a_1a_2a_3} \equiv D_b \pi^{a_1a_2a_3} - \tfrac{3}{2}\, f_{bc}{}^{[a_1|}\, \pi^{c|a_2a_3]}$\,. 
At least when $f_{ab}{}^c=0$\,, we have $\pi^{ijk}=\delta^i_a\delta^j_b\delta^k_c\,\pi^{abc}$ and the above properties are precisely the conditions for the tri-vector field to be a Nambu--Poisson structure. 
The hexa-vector $\pi^{i_1\cdots i_6}$ and similar poly-vector fields in Type IIB theory also satisfy similar properties.
It is suggestive to refer to these poly-vector fields as to the generalised Nambu--Poisson structures.

\subsubsection{Conventions}

In what follows we study concrete examples of non-abelian U-duality, for which the following ordering for various tensors will be used. 
When studying explicit examples of dualities the following dictionary ordering for various tensors is used. When considering the E${}_{5(5)}$ EDA in the M-theory picture, the generators $T_A=\bigl(T_a,\,\frac{1}{\sqrt{2!}}T^{a_1a_2},\,\frac{1}{\sqrt{5!}}T^{a_1\cdots a_5}\bigr)$ are ordered as
\begin{align}
 T_A = (& T_1,T_2,T_3,T_4,T_5,\nn\\
 & T^{12},T^{13},T^{14},T^{15},T^{23},T^{24},T^{25},T^{34},T^{35},T^{45},\nn\\
 & T^{12345})\,,
\end{align}
and various tensors, such as $\hat{\cM}_{AB}$\,, when written as matrices respect this ordering. 
In the Type IIB picture, the generators $T_A=\bigl(T_{\sfa},\,T_\alpha^{\sfa},\,\frac{1}{\sqrt{3!}}T^{\sfa_1\sfa_2\sfa_3}\bigr)$ are ordered as
\begin{align}
 T_A = (& T_1,T_2,T_3,T_4,\nn\\
 &T_{\mathbf{1}}^1,T_{\mathbf{1}}^2,T_{\mathbf{1}}^3,T_{\mathbf{1}}^4,T_{\mathbf{2}}^1,T_{\mathbf{2}}^2,T_{\mathbf{2}}^3,T_{\mathbf{2}}^4,\nn\\
 &T^{123},T^{124},T^{134},T^{234})\,.
\end{align}

\subsubsection{Spacetime signature}

Important for some of the examples considered in the text is the choice of the space-time signature. If the constant matrix $\hat{\cM}_{AB}$ is positive definite (and is an element of $\text{GL}(n)\subset \text{E}_{n(n)}$), it can be parameterized as
\begin{equation}
\label{eq:metparam}
 \hat{\cM}_{AB} = (\det\hat{g})^{\frac{1}{9-n}}  \begin{pmatrix}
 \hat{g}_{ab} & 0 & 0 \\
 0 & \!\! \hat{g}^{a_1a_2,b_1b_2} \!\! & 0 \\
 0 & 0 & \! \hat{g}^{a_1\cdots a_5,b_1\cdots b_5} \! 
\end{pmatrix} \! .
\end{equation}
For the examples we present it is required that $\hat{g}_{ab}$ has negative eigenvalues in order to satisfy the equations of motion. Let us comment more on this issue.\footnote{See \cite{Hull:1998vg,Hull:1998ym,Keurentjes:2004bv,Keurentjes:2004xx,deBuyl:2005it,Cook:2005wj,Malek:2013sp,Blair:2013gqa} for more details on the generalised metric with the Lorentzian signature.} 

Consider first two simple examples. For the first one, suppose that $\hat{\cM}_{AB}$ is given by
\begin{align}
 \hat{\cM}_{AB} = {\begin{pmatrix}
 -1 & 0 & 0 & 0 & 0 & 0 & 0 & 0 & 0 & 0 \\
  0 & 1 & 0 & 0 & 0 & 0 & 0 & 0 & 0 & 0 \\
  0 & 0 & 1 & 0 & 0 & 0 & 0 & 0 & 0 & 0 \\
  0 & 0 & 0 & 1 & 0 & 0 & 0 & 0 & 0 & 0 \\
  0 & 0 & 0 & 0 & 1 & 0 & 0 & 0 & 0 & 0 \\
  0 & 0 & 0 & 0 & 0 & 1 & 0 & 0 & 0 & 0 \\
  0 & 0 & 0 & 0 & 0 & 0 & 1 & 0 & 0 & 0 \\
  0 & 0 & 0 & 0 & 0 & 0 & 0 & -1 & 0 & 0 \\
  0 & 0 & 0 & 0 & 0 & 0 & 0 & 0 & -1 & 0 \\
  0 & 0 & 0 & 0 & 0 & 0 & 0 & 0 & 0 & -1\end{pmatrix}},
\end{align}
which is an element of $\text{SL}(5)$, however with negative eigenvalues. In this case, we need to modify the parameterization \eqref{eq:metparam} as
\begin{align}
 \hat{\cM}_{AB} = \abs{\det\hat{g}}^{\frac{1}{9-n}} \begin{pmatrix}
 \hat{g}_{ab} & 0 & 0 \\
 0 & \sigma\,\hat{g}^{a_1a_2,b_1b_2} & 0 \\
 0 & 0 & \hat{g}^{a_1\cdots a_5,b_1\cdots b_5} 
\end{pmatrix} ,
\label{eq:M-param}
\end{align}
where $\abs{\det\hat{g}}$ denotes the absolute value of $\det\hat{g}_{ab}$ and $\sigma$ should be identified as $\sigma=-1$\,. 
The internal metric then is possible to choose as $\hat{g}_{ab}=\diag(+1,-1,-1,-1)$\,. 
Another example is provided by the following choice of the generalised metric
\begin{align}
 \hat{\cM}_{AB} = {\begin{pmatrix}
  1 & 0 & 0 & 0 & 0 & 0 & 0 & 0 & 0 & 0 \\
  0 & -1 & 0 & 0 & 0 & 0 & 0 & 0 & 0 & 0 \\
  0 & 0 & -1 & 0 & 0 & 0 & 0 & 0 & 0 & 0 \\
  0 & 0 & 0 & -1 & 0 & 0 & 0 & 0 & 0 & 0 \\
  0 & 0 & 0 & 0 & 1 & 0 & 0 & 0 & 0 & 0 \\
  0 & 0 & 0 & 0 & 0 & 1 & 0 & 0 & 0 & 0 \\
  0 & 0 & 0 & 0 & 0 & 0 & 1 & 0 & 0 & 0 \\
  0 & 0 & 0 & 0 & 0 & 0 & 0 & -1 & 0 & 0 \\
  0 & 0 & 0 & 0 & 0 & 0 & 0 & 0 & -1 & 0 \\
  0 & 0 & 0 & 0 & 0 & 0 & 0 & 0 & 0 & -1\end{pmatrix}} \,,
\end{align}
which is also an element of $\text{SL}(5)$\,. This element is not related to the previous one via an $\text{SL}(5)$ transformation as the traces are different. 
For this example, we find $\hat{g}_{ab}=\diag(-1,+1,+1,+1)$ and again the sign should be $\sigma=-1$\,. 
That is to say, when $\hat{\cM}_{AB}$ is not positive definite, the parameterization must includes additional negative signs. The same prescription has been used in \cite{Bakhmatov:2020kul} to define tri-vector deformations of the AdS part of AdS${}_4\times \SS^7$ background.

Given the sign $\s$ in the parametrisation of $\hat{\cM}_{AB}$\,, the parameterization of the generalised metric $\hat{\cM}_{IJ}$ in \eqref{eq:gen-met} also must be modified. For example, for the M-theory section one writes
\begin{align}
 \hat{\cM}_{IJ} &= \abs{\det g}^{\frac{1}{9-n}} \begin{pmatrix}
 g_{ij} & 0 & 0 \\
 0 & \!\! \sigma\,g^{i_1i_2,j_1j_2} \!\! & 0 \\
 0 & 0 & g^{i_1\cdots i_5,j_1\cdots j_5} 
\label{eq:gen-met-mod}
\end{pmatrix}\! .
\end{align}
The presence of the factor $\s$ has curious effects when reducing from exceptional field theory formulation to that of standard supergravity. For $\s=+ 1$ the equations of motion of exceptional field theory are known to reduce to those of the standard supergravity (see for example \cite{Berman:2011jh}), while for $\sigma=-1$, the kinetic term in the 3-form potential has the wrong sign.
As it has been studied in \cite{Hull:1998ym} M-theory (or Type IIA theory) and the M$^*$-theory (or Type IIA$^*$ theory) are related by a timelike T-duality, and indeed in our examples, the dualised backgrounds are solutions of M$^*$-theory.
M$^*$-theory has two time directions and precisely the required wrong sign in the kinetic term of $C_3$\,. Hence, our examples with $\sigma=-1$ can be naturally interpreted as non-abelian generalisations of timelike U-dualities.\footnote{M'-theory with more time directions is also in the same orbit, although we do not consider such examples.} 
Similarly, in some of the Type IIB backgrounds, $m_{\alpha\beta}$ have negative determinant $\det m=-1$ and must be parameterized rather as
\begin{align}
 m_{\alpha\beta} = \begin{pmatrix} 1 & -C_0 \\ 0 & 1 \end{pmatrix}
 \begin{pmatrix} \Exp{-\Phi} & 0 \\ 0 & \sigma \Exp{\Phi} \end{pmatrix}
 \begin{pmatrix} 1 & 0 \\ -C_0 & 1 \end{pmatrix},
\label{eq:IIB*}
\end{align}
where $\sigma=-1$\,. 
This parameterizes a coset SL(2)/SO(1,1)\,. 
Given this sign, the kinetic term for $C_0$ has the wrong sign and ends up in Type IIB$^*$ theory.

\section{Examples of non-abelian U-duality}
\label{sec:Uexmpls}

In this section, we present a set of explicit examples of non-abelian U-duality (see \cite{Sakatani:2019zrs,Malek:2019xrf,Blair:2020ndg} for some discussion on simple examples). 
In particular, the following three types of examples are studied:
\begin{enumerate}
 \item[$\bullet$ A.] \underline{Uplifts of PL T-duality and small extensions}\\[1mm]
Here we study several uplifts of PL T-dualities ($T_a\leftrightarrow T^a$) in type IIA theory to M-theory. 
Since this duality contains a timelike T-duality, in all of the examples, a solution of M-theory $\sigma=+1$ is mapped to a solution of M$^*$-theory ($\sigma=-1$). 
In some examples due to the presence of non-trivial structure constants (denoted as $f_{\dot{a}\dot{b}}{}^5$), our examples go beyond the usual uplifts of PL T-dualities. 

 \item[$\bullet$ B.] \underline{Dualities between M-theory and Type IIB theory}\\[1mm]
We also study non-abelian extension of factorized T-dualities that connect M-theory solutions and Type IIB solutions. 
Again, the non-abelian dualities involve the timelike T-duality, and we observe the signature changes $\sigma=\pm 1\to \mp 1$\,.  

 \item[$\bullet$ C.] \underline{Generalized Yang--Baxter deformation}\\[1mm]
Generalized Yang--Baxter deformation is a class of non-abelian dualities that provides continuous deformations of the original geometry. 
Because of the continuity, the sign $\sigma$ is not changed under this type of deformations. 
We consider examples of generalised Yang--Baxter deformation both in M-theory and Type IIB theory.
These examples are not uplifts of the usual Yang--Baxter deformation (that corresponds to a local O($d,d$) transformation). 
\end{enumerate}
Under these three types of dualities, non-geometric backgrounds, such as non-Riemannian backgrounds \cite{Lee:2013hma} or U-folds, can be produced. 
A general procedure to identify the global structure of the dual geometries has not been developed, but a dual geometry generally contains non-vanishing (globally) non-geometric fluxes.
Then, one can regard the dual geometry as a U-fold by making a certain periodic identification. 
The explicit form of the monodromy matrix is given for each example. 
To express the monodromy matrices, we use the E${}_{n(n)}$ generators such as $R_{a_1a_2a_3}$ and $R_{a_1\cdots a_6}$ (see \cite{Sakatani:2020wah} for their explicit form). 

\subsection{Poisson--Lie T-duality and extensions}

Important property of Drinfel'd Lie algebra relevant for the formulation of PL T-duality/plurality is its symmetry under the exchange $T_a\leftrightarrow T^a$\,. 
For this reason any background has a canonical dual generated by the canonical matrix $C_A{}^B$ given in \eqref{eq:C-PL-T}. This matrix is simply a set of (formal) abelian T-dualities along all directions.

Evidently, for a general EDA, dimensions of the isotropic subalgebra $\gg$ spanned by generators $T_a$ and of its linear complement in the algebra $\tgg$ spanned by the rest generators
$\{T^{a_1a_2}, T^{a_1\dots a_5}\}$ do not match and one does not have a canonical map, such as $T_a\leftrightarrow T^{ab}$\,. A possible generalisation of the relation between isotropic subalgebra and its dual to the case of 11-dimensional backgrounds has been suggested in \cite{Musaev:2020bwm}, that is based on introducing an external automorphism of the corresponding abelian U-duality group, acting on representations of its maximal GL($n$) subgroup in the way, similar to the set of abelian T-dualities in all directions. This apparently does not cover all possible choices of the matrix $C$, only generalising non-abelian T-duality transformation, rather than the full set of Poisson--Lie-type symmetries. Hence, due to the lack of a general procedure, we are forced to find a non-trivial matrix $C_A{}^B$ for our examples on a case-by-case basis.

Before proceeding with explicit examples, let us elaborate more on the type of transformation discussed in \cite{Musaev:2020bwm}. For that, we restrict the discussion to such exceptional Drinfel'd algebras that are always mapped into another EDA under the PL T-duality \eqref{eq:C-PL-T}, i.e., T-dualities in all directions. 
Given the relation between double field theory and exceptional field theory, generators $T^{\dot{a}}$ ($\dot{a}=1,\dotsc,n-1$) in the Drinfel'd algebra should be identified with $T^{\dot{a}\#}$ ($\#$ corresponds to the M-theory direction). 
Therefore PL T-duality corresponds to
\begin{align}
 T'_{\dot{a}} = T^{{\dot{a}}\#}\,,\qquad T'^{{\dot{a}}\#} = T_{\dot{a}}\,.
\end{align}
For simplicity, we consider the case Type IIA theory is mapped Type IIA theory, and then $n$ should be odd. 
Then, for $n=3,5,7$, the ``PL T-duality'' for the EDA generators can be defined as follows:
\begin{align}
 &\underline{n=3:} &&\left\{
 \begin{aligned}
 &T'_{\dot{a}} = T^{\dot{a}3}\,,\\
 & T'_3 = T^{12}\,,\\
 &  T'^{12} = -T_3\,,\\
 &T'^{\dot{a}3} = T_{\dot{a}} \,,
 \end{aligned}
 \right.
\label{eq:transf-E3}
\\
 &\underline{n=5:}&&
 \left\{\begin{aligned}
 &T'_{a} = T^{a5}\,,\\
 &T'_5 = - T^{1\cdots 5}\,,
\\
 &T'^{ab} = -\tfrac{1}{2!}\,\epsilon_{abcd5}\,T^{cd}\,,\\
 & T'^{a5} = T_a\,,\\
 &
 T'^{1\cdots 5} = -T_5\,,
 \end{aligned}\right.
\label{eq:transf}
\\
 &\underline{n=7:}&&
 \left\{\begin{aligned}
 &T'_{\dot{a}} = T^{\dot{a}7}\,,\\
 & T'_5 = -T^{1\cdots7,7}\,,\\
 & T'^{\dot{a}_1\dot{a}_2} = -\tfrac{1}{5!}\,\epsilon_{\dot{a}_1\dot{a}_2b_1\cdots b_5}\,T^{b_1\cdots b_5}\,,\\
 &
 T'^{\dot{a} 7} = T_{\dot{a}}\,,
\\
 &T'^{\dot{a}_1\cdots\dot{a}_5} = \epsilon_{\dot{a}_1\cdots\dot{a}_5\dot{b} 7}\,T^{1\cdots7,\dot{b}}\,,\\
 &T'^{\dot{a}_1\cdots\dot{a}_47} = \tfrac{1}{2!}\,\epsilon_{\dot{a}_1\cdots\dot{a}_4\dot{b}_1\dot{b}_27}\,T^{\dot{b}_1\dot{b}_2}\,,
\\
 &T'^{1\cdots7,\dot{a}} = - \tfrac{1}{5!}\,\epsilon_{\dot{a}\dot{b}_1\cdots\dot{b}_5 7}\, T'^{\dot{b}_1\cdots\dot{b}_5}\,,\\
 & T'^{1\cdots7,7}= T_5 \,,
\end{aligned}\right.
\end{align}
where $\epsilon_{1\cdots n}=1$\,. 
The condition that such a transformation is indeed a symmetry of two EDA's (i.e.~relates two realisations of the same EDA) is highly restrictive. 
In the next sections we identify the constraints it imposes on structure constants for $n=3$ and $n=5$.

\subsubsection{\texorpdfstring{E${}_{3(3)}$}{E3} EDA}

We start with $n=3$, for which the notation E${}_{3(3)}$ stands for the SL(2)$\times$SL(3) group. In this case one finds that for the EDA to be dualised into an EDA the structure constants should satisfy
\begin{align}
 f_{12}{}^3 &= f_{13}{}^2 = f_{23}{}^1 = 0\,,\\
 f_{13}{}^3 &= f_{12}{}^2\,,\quad
 f_{23}{}^3 = f_{21}{}^1\,,\quad
 f_{32}{}^2 = f_{31}{}^1\,,\\
 Z_a &= \tfrac{1}{6}f_{ab}{}^b\,.
\end{align}
Given these constraints one is left with only six independent structure constants:
\begin{align}
 m_1 = f_{12}{}^1\,,\quad 
 m_2 = f_{12}{}^2\,,\quad 
 m_3 = f_{13}{}^1\,,\quad 
 n_a = - f_a{}^{123}\,,
\end{align}
and the Leibniz identity does not impose any further conditions. 
The map \eqref{eq:transf-E3} then corresponds to the folllwing exchange of the structure constants
\begin{align}
 m_a \leftrightarrow n_a \,.
\label{eq:ma-na}
\end{align}
which is similar to the PL T-duality.
Assuming $m_a=0$ one notices that this EDA contains a four-dimensional Drinfel'd double (with $f_{\dot{a}\dot{b}}{}^{\dot{c}}=0$ and $\tilde{f}^{ab}{}_c = -f_c{}^{ab3}$) as the subalgebra generated by $\{T_{\dot{a}},\,T^{\dot{a}3}\}$.
However, after the duality \eqref{eq:ma-na} $Z_a$ is produced and the generators $\{T_{\dot{a}},\,T^{\dot{a}3}\}$ does not satisfy the Drinfel'd algebra. Hence, the important observation is that the duality \eqref{eq:ma-na} \emph{is not simply an uplift of the PL T-duality}. 
Moreover, for $n=3$ one cannot realise PL T-duality as a non-abelian U-duality.

In order to generate solutions of ungauged supergravities, we expect that the unimodularity condition $X_{AB}{}^B=0$ should be satisfied. 
However, unfortunately, the unimodularity is equivalent to $m_a=n_a=0$ and there are no non-trivial examples here. 

\subsubsection{\texorpdfstring{E${}_{5(5)}$}{E5} EDA}

More fruitful is the case $n=5$ for which the notation E${}_{5(5)}$ stands for the SO(5,5) group. We start by a class of EDA's given by 
\begin{align}
 f_{\dot{a}\dot{b}}{}^{\dot{b}} = f_{a5}{}^{b} = 0\,,\quad
 f_{\dot{b}}{}^{\dot{b}\dot{a}5} = f_{\dot{a}}{}^{\dot{b}\dot{c}\dot{d}} = f_5{}^{abc} = 0\,,\quad
 Z_a =0\,,
\end{align}
and find that the transformation \eqref{eq:transf} always maps an EDA to an EDA. 
For the remaining non-vanishing structure constants one reads off the transformation rules from \eqref{eq:transf} as:
\begin{equation}
\label{eq:PL-T-5}
\begin{aligned}
 f'_{\dot{a}\dot{b}}{}^{\dot{c}} &= - f_{\dot{c}}{}^{\dot{a}\dot{b}5}\,,\qquad
 f'_{\dot{a}}{}^{\dot{b}\dot{c}5} = - f_{\dot{b}\dot{c}}{}^{\dot{a}}\,,\\
 f'_{\dot{a}\dot{b}}{}^5 &= - \tfrac{1}{2!}\,\epsilon_{\dot{a}\dot{b}\dot{c}\dot{d}5}\, f_{\dot{c}\dot{d}}{}^5 \,.
\end{aligned}
\end{equation}
The first two relations are the same as that for PL T-duality under the identification $f_{\dot{a}}{}^{\dot{b}\dot{c}} = -f_{\dot{a}}{}^{\dot{b}\dot{c}5}$\,. 
The last one however is specific to EDA and hence again makes the transformation beyond a simple uplift. For that, one would consider $f_{\dot{a}\dot{b}}{}^5=0$\, and obtain a non-abelian U-duality uplift of the PL T-duality.

Let us now present several examples of non-abelian U-dualities with both vanishing and non-vanishing $f_{\dot{a}\dot{b}}{}^5$\,. 

\medskip

\paragraph{Example 1 ($f_{\dot{a}\dot{b}}{}^5=0$)\\[2mm]}

Consider an EDA generated by
\begin{align}
 f_{12}{}^3 = -1\,,\quad
 f_{14}{}^2 = 1\,,\quad
 f_{24}{}^1 = -1\,.
\end{align}
Group element parameterization is chosen as
\begin{align}
 g= \Exp{x\,T_1 + y\,T_2+z\,T_3+u\,T_5} \Exp{w\,T_4} ,
\end{align}
that gives the following right-invariant 1-forms 
\begin{align}
 r_i{}^a = \begin{pmatrix}
 1 & 0 & \frac{y}{2} & 0 & 0 \\
 0 & 1 & -\frac{x}{2} & 0 & 0 \\
 0 & 0 & 1 & 0 & 0 \\
 -y & x & -\frac{x^2+y^2}{2} & 1& 0 \\
 0 & 0 & 0 & 0 & 1
 \end{pmatrix}.
\end{align}
The flat metric and the metric $g_{ij} \equiv r_i^a\,r_j^b\,\hat{g}_{ab}$ are then
\begin{equation}
\begin{aligned}
 \!\!\!\! \hat{g}_{ab} = \begin{pmatrix}
 4 & 0 & 0 & 0 & 0 \\
 0 & 2 & 0 & 0 & 0 \\
 0 & 0 & 0 & 2 & 0 \\
 0 & 0 & 2 & 0 & 0 \\
 0 & 0 & 0 & 0 & 1
 \end{pmatrix}\! , \ \ 
 g_{ij} =\begin{pmatrix}
 4 & 0 & 0 & -3 y & 0 \\
 0 & 2 & 0 & x & 0 \\
 0 & 0 & 0 & 2 & 0 \\
 \!\! -3 y & x & 2 & 2 y^2 & 0 \\
 0 & 0 & 0 & 0 & 1
\end{pmatrix}\! .
\end{aligned}
\end{equation}
Taking for the external metric simply the unity matrix $g_{\mu\nu} = \diag(1,1,1,1,1,1)$\,, one obtains a vacuum solution of 11-dimensional supergravity. 
The duality-invariant external metric is then found to be
\begin{align}
 \mathfrak{g}_{\mu\nu} \equiv \abs{\det g_{ij}}^{1/4}\,g_{\mu\nu} = 2^{5/4}\,\diag(1,1,1,1,1,1)\,.
\end{align}
Considering the $x^5$ to be the M-theory direction this is reduced to the following Type IIA solution
\begin{align}
 g_{ij} = \begin{pmatrix}
 4 & 0 & 0 & -3 y \\
 0 & 2 & 0 & x \\
 0 & 0 & 0 & 2 \\
 -3 y & x & 2 & 2 y^2 
\end{pmatrix},\quad g_{\mu\nu} = \diag(1,1,1,1,1,1). 
\label{eq:NATD-1}
\end{align}
Given the signature of the background, one constructs the constant matrix $\hat{\cM}_{AB}\in \text{E}_{5(5)}$ with $\sigma=+1$\,. 

Perform now the transformation \eqref{eq:PL-T-5}, that gives
\begin{align}
 f'_3{}^{12 5} = 1\,,\quad
 f'_2{}^{14 5} = -1\,,\quad
 f'_1{}^{24 5} = 1\,.
\end{align}
Parameterizing the element of the dual subgroup \\ $g'=\exp[x'\,T'_1+y'\,T'_2+z'\,T'_3+u'\,T'_4+v'\,T'_5]$ one finds the right-invariant 1-forms trivial and the Nambu--Lie structure as follows
\begin{equation}
\begin{aligned}
 \pi'& = \tfrac{1}{3!}\,x'^a\,f'_a{}^{bcd}\,\partial_b\wedge\partial_c\wedge\partial_d\\
& = \bigl(z'\,\partial_1\wedge\partial_2 -y' \,\partial_1\wedge\partial_4 + x'\,\partial_2\wedge\partial_4\bigr)\wedge\partial_5\,.
\end{aligned}
\end{equation}

Construction of the generalised frame fields $E'_A{}^I$ and the generalised metric $ \cM'_{IJ} = E'_I{}^A\,E'_J{}^B\,\hat{\cM}'_{AB}\,$ is then straightforward. However, one finds, that to recover a solution to supergravity equations of motion, one has to use the parameterization with $\sigma=-1$\,. 
The dual internal metric and the 3-form potential then become
\begin{align}
 g'_{ij} &= \frac{2^{\frac{5}{3}}}{(8 + z'^2)^{\frac{2}{3}}}
\begin{pmatrix}
 1 & 0 & \frac{x' z'}{4} & 0 & 0 \\
 0 & 2 & \frac{y' z'}{4} & 0 & 0 \\
 \frac{x' z'}{4} & \frac{y' z'}{4} & -\frac{2 x'^2+y'^2}{4} & \frac{8+z'^2}{4} & 0 \\
 0 & 0 & \frac{8+z'^2}{4} & 0 & 0 \\
 0 & 0 & 0 & 0 & -\frac{1}{8} 
\end{pmatrix}, 
\\
 C'_3 &= \frac{\bigl(z'\,\rmd x' \wedge \rmd y' - y'\, \rmd x' \wedge \rmd z' + 2\,x'\,\rmd y' \wedge \rmd z'\bigr) \wedge \rmd v'}{8+z'^2}\,.\nn
\end{align}
 Assuming invariance of the (abelian) U-duality-invariant metric  $\mathfrak{g}'_{\mu\nu}$ under such non-abelian transformations, the external metric is found as
\begin{align}
 g'_{\mu\nu} = 2^{\frac{2}{3}} (8 + z'^2)^{\frac{1}{3}}\,\diag(1,1,1,1,1,1)\,.
\end{align}
Hence, one obtains a background of M$^*$-theory, concluding that in this case the transformation acts as a timelike U-duality. 

Indeed, reducing to 10-dimensions one ends up with a background of the Type IIA$^*$ theory:
\begin{align}
 g'_{mn} &= \frac{2}{8 + z'^2}
\begin{pmatrix}
 1 & 0 & \frac{x' z'}{4} & 0 \\
 0 & 2 & \frac{y' z'}{4} & 0 \\
 \frac{x' z'}{4} & \frac{y' z'}{4} & -\frac{2 x'^2+y'^2}{4} & \frac{8+z'^2}{4} \\
 0 & 0 & \frac{8+z'^2}{4} & 0 
\end{pmatrix},
\nn\\
 g'_{\mu\nu} &= 2^{\frac{2}{3}} (8 + z'^2)^{\frac{1}{3}}\,\diag(1,\dotsc,1)\,,
\label{eq:NATD-2}
\\
 B'_2 &= \frac{z'\,\rmd x' \wedge \rmd y' - y'\, \rmd x' \wedge \rmd z' + 2\,x'\,\rmd y' \wedge \rmd z'}{8+z'^2}\,,
\nn\\
 \Exp{-2\Phi'} &= 4\,(8 + z'^2)\,.\nn
\end{align}
Explicit check shows that the map between \eqref{eq:NATD-1} and \eqref{eq:NATD-2} is precisely the standard non-abelian T-duality. 
Hence, this example is an 11-dimensional uplift of non-abelian T-duality. 

We also note that the generalised metric satisfies
\begin{align}
\begin{split}
 &\cM'_{IJ}(x'+c^x,y'+c^y,z') 
\\
 &= \bigl(\Omega_{c^x}\,\Omega_{c^y}\,\cM'\,\Omega^{\rm T}_{c^y}\,\Omega^{\rm T}_{c^x}\bigr)_{IJ}(x',y',z')\,,
\end{split}
\end{align}
where
\begin{align}
 \Omega_{c^x}\equiv \Exp{c^x\,R_{245}}\in \text{E}_{5(5)},\quad \Omega_{c^y}\equiv\Exp{-c^y\,R_{145}} \in \text{E}_{5(5)} .
\end{align}
Therefore, if we identify the coordinates as $x'\sim x'+c^x$ and $y'\sim y'+c^y$\,, we get a U-fold. 
If we regard $x^5$ as the M-theory direction, $R_{245}$ and $R_{145}$ are generators of the $\beta$-transformation, and this background corresponds to a 11-dimensional uplift of a T-fold. 

\medskip

\paragraph{Example 2 ($f_{\dot{a}\dot{b}}{}^5\neq 0$)\\[2mm]}

Starting from precisely the same background as in the previous example one is able to arrive at a non-Riemannian background. For that it is convenient to rearrange generators as
\begin{align}
 f_{12}{}^5 = -1\,,\quad
 f_{13}{}^2 = 1\,,\quad
 f_{23}{}^1 = -1\,.
\end{align}
Explicit check shows that for such defined structure constants the following choice of the generalised metric $\hat{\cM}_{AB}$ 
\begin{equation}
\hat{\cM}_{AB} = 2^{\frac{5}{4}}
{
\begin{pmatrix}
 4 & 0 & 0 & 0 & 0 & 0 & 0 & 0 & 0 & 0 & 0 & 0 & 0 & 0 & 0 & 0 \\
 0 & 2 & 0 & 0 & 0 & 0 & 0 & 0 & 0 & 0 & 0 & 0 & 0 & 0 & 0 & 0 \\
 0 & 0 & 0 & 0 & 2 & 0 & 0 & 0 & 0 & 0 & 0 & 0 & 0 & 0 & 0 & 0 \\
 0 & 0 & 0 & 1 & 0 & 0 & 0 & 0 & 0 & 0 & 0 & 0 & 0 & 0 & 0 & 0 \\
 0 & 0 & 2 & 0 & 0 & 0 & 0 & 0 & 0 & 0 & 0 & 0 & 0 & 0 & 0 & 0 \\
 0 & 0 & 0 & 0 & 0 & \frac{1}{8} & 0 & 0 & 0 & 0 & 0 & 0 & 0 & 0 & 0 & 0 \\
 0 & 0 & 0 & 0 & 0 & 0 & 0 & 0 & \frac{1}{8} & 0 & 0 & 0 & 0 & 0 & 0 & 0 \\
 0 & 0 & 0 & 0 & 0 & 0 & 0 & \frac{1}{4} & 0 & 0 & 0 & 0 & 0 & 0 & 0 & 0 \\
 0 & 0 & 0 & 0 & 0 & 0 & \frac{1}{8} & 0 & 0 & 0 & 0 & 0 & 0 & 0 & 0 & 0 \\
 0 & 0 & 0 & 0 & 0 & 0 & 0 & 0 & 0 & 0 & 0 & \frac{1}{4} & 0 & 0 & 0 & 0 \\
 0 & 0 & 0 & 0 & 0 & 0 & 0 & 0 & 0 & 0 & \frac{1}{2} & 0 & 0 & 0 & 0 & 0 \\
 0 & 0 & 0 & 0 & 0 & 0 & 0 & 0 & 0 & \frac{1}{4} & 0 & 0 & 0 & 0 & 0 & 0 \\
 0 & 0 & 0 & 0 & 0 & 0 & 0 & 0 & 0 & 0 & 0 & 0 & 0 & 0 & -\frac{1}{2} & 0 \\
 0 & 0 & 0 & 0 & 0 & 0 & 0 & 0 & 0 & 0 & 0 & 0 & 0 & -\frac{1}{4} & 0 & 0 \\
 0 & 0 & 0 & 0 & 0 & 0 & 0 & 0 & 0 & 0 & 0 & 0 & -\frac{1}{2} & 0 & 0 & 0 \\
 0 & 0 & 0 & 0 & 0 & 0 & 0 & 0 & 0 & 0 & 0 & 0 & 0 & 0 & 0 & -\frac{1}{32} 
 \end{pmatrix}}
\end{equation}
provides a solution to equations of motion of 11-dimensional supergravity.
The transformation \eqref{eq:PL-T-5} then renders
\begin{align}
 f'_{34}{}^5 = 1\,,\quad
 f'_2{}^{135} = -1\,,\quad
 f'_1{}^{235} = 1\,.
\end{align}
We find that the corresponding dual geometry becomes a non-Riemannian background for which the metric cannot be defined. See \cite{Cho:2018alk,Berman:2019izh,Park:2020ixf} for more discussion on such backgrounds. To cure that we introduce additional twist $\exp[\eta\,R_{123}]$ which gives
\begin{align}
 f'_{34}{}^5 = 1\,,\ \ 
 f'_2{}^{135} = -1\,,\ \ 
 f'_1{}^{235} = 1\,,\ \ 
 f'_4{}^{125} = -\eta\,.
\end{align}
Using the parametrization $g'=\exp[x'\,T'_1+y'\,T'_2+z'\,T'_3] \exp[u'\,T'_4+v'\,T'_5]$ we find the following right-invariant 1-form
\begin{align}
 r'_i{}^a = {\begin{pmatrix}
 1 & 0 & 0 & 0 & 0 \\
 0 & 1 & 0 & 0 & 0 \\
 0 & 0 & 1 & 0 & 0 \\
 0 & 0 & 0 & 1 & z' \\
 0 & 0 & 0 & 0 & 1
\end{pmatrix}},
\end{align}
and the Nambu--Lie structure
\begin{align}
 \pi' = \bigl(-y'\,\partial_1\wedge\partial_3 + x'\,\partial_2\wedge\partial_3 - \eta\,u'\,\partial_1\wedge\partial_2 \bigr)\wedge \partial_5\,.
\end{align}
Recovering the generalised metric $\cM_{IJ}$ one again finds that the sign choice must be $\sigma=-1$.
The dual 11-dimensional fields are then given by
\begin{align}
\begin{split}
 g_{ij} &= \tfrac{(2\eta)^{-\frac{4}{3}}}{(8 + u^2)^{\frac{2}{3}}}
{\begin{pmatrix}
 8-x^2 & -x y & \eta u x & \eta x z & \eta x \\
 -x y & 16-y^2 & \eta u y & \eta y z & \eta y \\
 \eta u x & \eta u y & \eta ^2 \left(-u^2\right) & -\eta (\eta u z+16) & -\eta^2 u \\
 \eta x z & \eta y z & -\eta (\eta u z+16) & \eta z (4 u-\eta z)-4 x^2-2 y^2+32 & \eta (2 u-\eta z) \\
 \eta x & \eta y & -\eta^2 u & \eta (2 u-\eta z) & -\eta^2 
\end{pmatrix}},
\\
 C_3 &=\frac{1}{\eta \,(8+u^2)}\bigg[8\,\rmd x\wedge\rmd y\wedge\rmd z - \bigl[\bigl(\tfrac{16-2 x^2-y^2}{\eta}+u z\bigr)\,\rmd x\wedge\rmd y + (u y\,\rmd x - 2 u x\,\rmd y)\wedge \rmd z\bigr]\wedge\rmd u\bigg]
\\
 &\quad - \frac{\bigl(u\,\rmd x\wedge \rmd y - y\,\rmd x\wedge\rmd u + 2 x\,\rmd y\wedge\rmd u \bigr)\wedge \rmd v}{\eta\,(8+u^2)}\,,\\
 g_{\mu\nu} &= 2^{-\frac{7}{12}}\,\eta^{\frac{2}{3}}\,\big(8+u^2\big)^{\frac{1}{3}}\,\diag(1,1,1,1,1,1)\,.
\end{split}
\end{align}
These satisfy equations of motion on backgrounds of M$^*$-theory. 
In the limit $\eta\to 0$, the supergravity fields become singular and we get a non-Riemannian background. 
Since the equations of motion are satisfied for arbitrary value of $\eta$\,, we expect that the non-Riemannian background also satisfies the equations of motion of the E${}_{5(5)}$ exceptional field theory. 

If we regard $x^5=v$ as the coordinate on the M-theory circle, this background gives the RR 1-form and 3-form fields although the original background was purely gravitational. 
This shows that this example also stands beyond the scope of the usual non-abelian T-duality, which does not generate RR fields.

The generalised metric in the dual geometry satisfies
\begin{align}
\begin{split}
 &\cM_{IJ}(x+c^x,y+c^y,u) 
\\
 &= \bigl(\Omega_{c^x}\,\Omega_{c^y}\,\cM'\,\Omega^{\rm T}_{c^y}\,\Omega^{\rm T}_{c^x}\bigr)_{IJ}(x,y,u)\,.
\end{split}
\end{align}
where
\begin{align}
 \Omega_{c^x}= \Exp{c^x\,R_{235}}\in \text{E}_{5(5)},\quad \Omega_{c^y}=\Exp{-c^y\,R_{135}} \in \text{E}_{5(5)}\,.
\end{align}
If we identify the coordinates as $x\sim x+c^x$ and $y\sim y+c^y$\,, we again obtain a U-fold, which is a 11-dimensional uplift of a T-fold. 

\medskip

\paragraph{Example 3 ($f_{\dot{a}\dot{b}}{}^5\neq 0$)\\[2mm]}

Another example with non-vanishing $f_{\dot{a}\dot{b}}{}^5$ is provided by the following Lie algebra
\begin{align}
 f_{23}{}^1 =1\,,\quad f_{34}{}^5=1\,,\quad f_{24}{}^3=c_0 \qquad (\abs{c_0}<1)\,.
\end{align}
Using the parameterization $ g= \Exp{x\,T_1}\Exp{y\,T_2}\Exp{z\,T_3}\Exp{u\,T_4}\Exp{v\,T_5}$ 
we obtain the right-invariant 1-form
\begin{align}
 r_i{}^a ={\begin{pmatrix}
 1 & 0 & 0 & 0 & 0 \\
 0 & 1 & 0 & 0 & 0 \\
 y & 0 & 1 & 0 & 0 \\
 \frac{c_0\,y^2}{2} & 0 & c_0\,y & 1 & z \\
 0 & 0 & 0 & 0 & 1\end{pmatrix}}.
\end{align}
Then the flat metric and the internal metric take the following form
\begin{equation}
\begin{aligned}
 \hat{g}_{ab}&= {\begin{pmatrix}
 0 & 0 & 0 & 1 & 0 \\
 0 & 1-c_0^2 & 0 & 0 & 0 \\
 0 & 0 & 1 & 0 & 0 \\
 1 & 0 & 0 & 0 & 0 \\
 0 & 0 & 0 & 0 & 1\end{pmatrix}},\\
 g_{ij} &= {\small\begin{pmatrix}
 0 & 0 & 0 & 1 & 0 \\
 0 & 1-c_0^2 & 0 & 0 & 0 \\
 0 & 0 & 1 & (1+c_0)\,y & 0 \\
 1 & 0 & (1+c_0)\,y & c_0\,(1+c_0)\,y^2+z^2 & z \\
 0 & 0 & 0 & z & 1
\end{pmatrix}}.
\end{aligned}
\end{equation}
Taking the external metric to be the unit matrix $g_{\mu\nu}=\diag(1,1,1,1,1,1)$ as before, one obtains a solution of 11-dimensional supergravity. Signature choice here is $\sigma=+1$. 

Under the transformation \eqref{eq:PL-T-5}, the structure constants get mapped into
\begin{align}
 f'_{12}{}^5 = -1 \,,\qquad f'_1{}^{235}=-1\,,\qquad f'_3{}^{245}=-c_0\,.
\end{align}
To recover the corresponding right-invariant 1-form one uses the $g'=\Exp{x'\,T'_1+y'\,T'_2+z'\,T'_3+u'\,T'_4+v'\,T'_5}$ for the dual group element and obtains
\begin{align}
 r'_i{}^a = {\begin{pmatrix}
 1 & 0 & 0 & 0 & \frac{y'}{2} \\
 0 & 1 & 0 & 0 & -\frac{x'}{2} \\
 0 & 0 & 1 & 0 & 0 \\
 0 & 0 & 0 & 1 & 0 \\
 0 & 0 & 0 & 0 & 1 
\end{pmatrix}}.
\end{align}
The Nambu--Lie structure is then
\begin{align}
 \pi' = \bigl(- x'\,\partial_2\wedge\partial_3 - c_0\,z'\,\partial_2\wedge\partial_4 \bigr)\wedge \partial_5\,.
\end{align}
To recover a solution from the corresponding generalised frame one must use a parameterization with $\sigma=-1$ for which the internal and external components of the metric and the 3-form potential become (the primes have been omitted)
\begin{align}
 g_{ij} &= \frac{1}{(1-c_0^2+x^2)^{\frac{2}{3}}}\times 
\nn\\
 &\quad \begin{pmatrix}
 -c_0^2 z^2-\frac{y^2}{4} & \frac{x y}{4} & -c_0 x z & 1-c_0^2+x^2 & -\frac{y}{2} \\
 \frac{x y}{4} & 1-\frac{x^2}{4} & 0 & 0 & \frac{x}{2} \\
 -c_0 x z & 0 & 1-c_0^2 & 0 & 0 \\
 1-c_0^2+x^2 & 0 & 0 & 0 & 0 \\
 -\frac{y}{2} & \frac{x}{2} & 0 & 0 & -1 
\end{pmatrix} ,
\nn\\
 C_3 &= \frac{1}{1-c_0^2+x^2}\,\bigl( \tfrac{xy}{2} \,\rmd x \wedge \rmd y \wedge \rmd z + x\,\rmd y \wedge \rmd z \wedge \rmd v 
\nn\\
 &\quad\qquad\qquad\qquad -c_0\,z\,\rmd x\wedge \rmd y\wedge \rmd v \bigr)\,,
\\
  g_{\mu\nu} &= \frac{(1 - c_0^2 + x^2)^{\frac{1}{3}}}{(1-c_0^2)^{\frac{1}{4}}}\,\diag(1,1,1,1,1,1)\,.\nn
\end{align}
We can check that they satisfy the equations of motion of M$^*$-theory.
This is also a non-trivial example of non-abelian U-duality that cannot be realised as a non-abelian T-duality.

Here, the dual geometry satisfies
\begin{align}
\begin{split}
 \cM_{IJ}(x,y,z+c^z) &= \bigl(\Omega_{c^z}\,\cM'\,\Omega^{\rm T}_{c^z}\bigr)_{IJ}(x,y,z)\,,
\\
 \Omega_{c^z}&=\Exp{-c_0\,c^z\,R_{245}} \in \text{E}_{5(5)}\,.
\end{split}
\end{align}
If we make an identification, $z\sim z+c^z$\,, we obtain a 11-dimensional uplift of a T-fold.

\subsection{Non-abelian duality between M-theory and Type IIB backgrounds}

In this section we present examples of non-abelian U-duality which connects M-theory backgrounds to Type IIB backgrounds. 

\subsubsection{\texorpdfstring{E${}_{3(3)}$}{E3} example}

To start with, let us consider the simplest non-trivial example, that is the E${}_{3(3)}$ EDA defined by the following choice of the structure constants
\begin{align}
 f_{23}{}^1 = 1 \,.
\end{align}
Parameterizing the group element as $g=\Exp{x\,T_1}\Exp{y\,T_2}\Exp{z\,T_3}$ one arrives at the following right-invariant 1-form 
\begin{align}
 r= \rmd x\,T_1 + \rmd y\,T_2 + \rmd z\,(T_3 + y\,T_1)\,.
\end{align}
The flat metric and the internal metric are given by
\begin{equation}
\begin{aligned}
 \hat{g}_{ab} = \begin{pmatrix} 0 & 0 & 1 \\ 0 & 1 & 0 \\ 1 & 0 & 0 \end{pmatrix}, &\quad &
 g_{ij} = \begin{pmatrix} 0 & 0 & 1 \\ 0 & 1 & 0 \\ 1 & 0 & 2\,y \end{pmatrix}.
\end{aligned}
\end{equation}
As before the external metric is chosen to be just a unity matrix and duality-invariant metric appears to be the same:
\begin{align}
 g_{\mu\nu} = \diag(1,1,1,1,1,1,1) = \mathfrak{g}_{\mu\nu}\,.
\end{align}
Such constructed 11-dimensional background is locally flat and is a trivial solution to supergravity equations of motion. Since the metric diverges as $y \to \infty$, it is more suggestive to thing of the $y$ direction as of a circle. In this case the U-dual background will be a U-fold with an SL(2) monodromy matrix (see below).

Consider now the flat metric $\hat{\cM}_{AB} \in \text{E}_{3(3)}\times \mathbb{R}^+$ and write it explicitly in the form 
\begin{align}
 \hat{\cM}_{AB}
 ={\begin{pmatrix}
  0 & 0 & 1 & 0 & 0 & 0 \\
  0 & 1 & 0 & 0 & 0 & 0 \\
  1 & 0 & 0 & 0 & 0 & 0 \\
  0 & 0 & 0 & 0 & 0 & -1 \\
  0 & 0 & 0 & 0 & -1 & 0 \\
  0 & 0 & 0 & -1 & 0 & 0
 \end{pmatrix}},
\end{align}
which corresponds to the parameterization \eqref{eq:M-param} with $\sigma=+1$\,. 
For the transformation matrix $T_A\to T'_A=C_A{}^B\,T_B$ we take
\begin{align}
 C_A{}^B = {\begin{pmatrix}
 0 & 0 & 0 & 0 & 1 & 0 \\
 0 & 1 & 0 & 0 & 0 & 0 \\
 1 & 0 & 0 & 0 & 0 & 0 \\
 0 & 0 & 0 & 0 & 0 & 1 \\
 0 & 0 & 1 & 0 & 0 & 0 \\
 0 & 0 & 0 & 1 & 0 & 0\end{pmatrix}}. 
\end{align}
If one regards $z$ as the M-theory direction, the above would correspond to the factorized T-duality along the $T_1$ direction.
The transformed generators $T'_A$ then satisfy the Type IIB EDA with
\begin{align}
 f'_{2\mathbf{2}}{}^{\mathbf{1}} = 1\,,
\end{align}
where $f'_{2\mathbf{2}}{}^{\mathbf{1}}$ denotes the transformed structure constant $f'_{a\alpha}{}^{\beta}$ with $\{a,\alpha,\beta\}=\{2,\mathbf{2},\mathbf{1}\}$\,. 
Parameterizing the dual the group element as $g'=\Exp{x'\,T'_1+y'\,T'_2}$ one obtains the following twist matrix
\begin{align}
 E'_I{}^A = {\begin{pmatrix}
 1 & 0 & 0 & 0 & 0 & 0 \\
 0 & 1 & 0 & 0 & 0 & 0 \\
 0 & 0 & 1 & 0 & 0 & 0 \\
 0 & 0 & 0 & 1 & 0 & 0 \\
 0 & 0 & y' & 0 & 1 & 0 \\
 0 & 0 & 0 & y' & 0 & 1 \end{pmatrix}}. 
\label{eq:SL2-twist}
\end{align}
The dual generalised metric becomes simply
\begin{align}
 \cM'_{IJ} = E'_I{}^A\,E'_J{}^B\,\cM'_{AB}
 = {\begin{pmatrix}
 \mathsf{g}_{mn} & 0 \\
 0 & m_{\alpha\beta}\,\mathsf{g}^{mn} \end{pmatrix}} 
\end{align}
with the (Einstein-frame) metric $\mathsf{g}_{mn}$ and $m_{\alpha\beta}$ given by
\begin{align}
 \mathsf{g}_{mn} =\begin{pmatrix} -1 & 0 \\ 0 & 1 \end{pmatrix},\qquad
 m_{\alpha\beta} =\begin{pmatrix} 0 & -1 \\ -1 & -2\,y \end{pmatrix}.
\end{align}
One observes, that $m_{\alpha\beta}$ has the determinant $-1$\,, which forces to parameterize this matrix as \eqref{eq:IIB*} with $\sigma=-1$\,, rendering the RR 0-form and the dilaton to be
\begin{align}
 C_0 = -\frac{1}{2\,y}\,,\qquad 
 \Exp{\Phi} = 2\,y\,.
\end{align}
The external metric (in the Einstein frame) is flat and finally for the 10-dimensional background we have 
\begin{equation}
\begin{aligned}
 g_{mn}&=\sqrt{2\,y}\,\diag(-1,1,\dotsc,1)\,,\qquad 
 C_0 = -\frac{1}{2\,y}\,,\\ 
 \Exp{\Phi}&= 2\,y\,,
\end{aligned}
\end{equation}
that is a solution of Type IIB$^*$ theory background equations of motion. Technically, the wrong sign in the kinetic term of $C_0$ comes from the minus sign in front of $\Exp{\Phi}$ in the parameterization of $m_{\alpha\beta}$\,. 

To arrive at solutions of more conventional supergravity theories one has to perform an abelian T-duality transformation along the timelike direction. This produces the following solution of Type IIA supergravity equations of motion
\begin{align}
 g_{mn}&= \diag\bigl(-\tfrac{1}{\sqrt{2\,y}},\sqrt{2\,y},\dotsc,\sqrt{2\,y}\bigr)\,,\qquad 
 C_1 = -\frac{\rmd t}{2\,y} \,,
\nn\\
 \Exp{-2\Phi} &= (2\,y)^{-\frac{3}{2}}\,.
\end{align}
Additionally performing abelian T-duality along the $x$-direction and further performing S-duality transformation, we obtain a non-trivial purely NS--NS background
\begin{equation}
\begin{aligned}
 \rmd s^2&= \frac{-\rmd t^2+\rmd x^2}{2\,y} + \rmd y^2 + \rmd z_1^2+\cdots + \rmd z_7^2 \,,\\
 B_2 &= \frac{\rmd t\wedge \rmd x}{2\,y} \,,\qquad 
 \Exp{-2\Phi} = 2\,y \,.
\end{aligned}
\end{equation}
One concludes, that using non-abelian U-duality transformation one is able to generate a non-trivial solution from a flat 10-dimensional Minkowski space.

The twist matrix \eqref{eq:SL2-twist} clearly shows that if we identify the coordinates as $y'\sim y'+c^y$\,, this background becomes a U-fold with the monodromy matrix given by the SL(2) S-duality transformation.

\subsubsection{\texorpdfstring{E${}_{5(5)}$}{E5} example}

To provide an example with $n=5$ let us start with structure constants \begin{align}
 f_{12}{}^2 = -1\,,\qquad 
 f_{13}{}^3 = 1\,,\qquad
 f_{23}{}^4 = -1\,,
\end{align}
and the right-invariant 1-form 
\begin{align}
 r = \rmd x\,T_1 + \Exp{-x}\rmd y\,T_2 + \rmd z\,\bigl(\Exp{x} T_3 -y\, T_4\bigr) + \rmd w \,T_4\,,
\end{align}
generated by the parametrization $ g= \Exp{x\,T_1}\Exp{y\,T_2+z\,T_3+w\,T_4},$ of the group element.
The flat and the internal metric read
\begin{equation}
\label{eq:M-hatg}
 \begin{aligned}
 \hat{g}_{ab} = \begin{pmatrix}
 1 & 0 & 0 & 2 \\
 0 & 1 & 0 & 0 \\
 0 & 0 & 1 & 0 \\
 2 & 0 & 0 & 0
\end{pmatrix} , &&
 g_{ij} = 
 \begin{pmatrix}
 1 & 0 & -2 y & 2 \\
 0 & \Exp{-2 x} & 0 & 0 \\
 -2 y & 0 & \Exp{2 x} & 0 \\
 2 & 0 & 0 & 0
 \end{pmatrix}.
\end{aligned}
\end{equation}
The external metric is again taken to be just the unity matrix $g_{\mu\nu} = \diag(1,1,1,1,1,1)$ rendering the duality-invariant metric $\mathfrak{g}_{\mu\nu} = \sqrt{2}\,\diag(1,1,1,1,1,1)$\,. 
This is an 11-dimensional Ricci flat space-time and apparently respects equations of motion of supergravity. 
When we construct the generalised metric, we choose the sign $\sigma=-1$ for convenience (since there is no 3-form potential, this geometry is a solution both of M-theory and M$^*$-theory). 

For duality transformation we take the following dualisation matrix
\begin{equation}
 C_A{}^B = 
 {\footnotesize\begin{pmatrix}
 1 & 0 & 0 & 0 & 0 & 0 & 0 & 0 & 0 & 0 & 0 & 0 & 0 & 0 & 0 & 0 \\
 0 & 1 & 0 & 0 & 0 & 0 & 0 & 0 & 0 & 0 & 0 & 0 & 0 & 0 & 0 & 0 \\
 0 & 0 & 0 & -1 & 0 & 0 & 0 & 0 & 0 & 0 & 0 & 0 & 0 & 0 & 0 & 0 \\
 0 & 0 & 0 & 0 & 0 & 0 & 1 & 0 & 0 & 0 & 0 & 0 & 0 & 0 & 0 & 0 \\
 0 & 0 & 0 & 0 & 0 & 0 & 0 & 0 & 0 & 0 & 0 & 1 & 0 & 0 & 0 & 0 \\
 0 & 0 & 0 & 0 & 0 & 0 & 0 & 0 & 0 & 0 & 0 & 0 & 0 & 0 & -1 & 0 \\
 0 & 0 & 0 & 0 & 0 & 0 & 0 & 0 & 0 & 0 & 0 & 0 & 0 & 1 & 0 & 0 \\
 0 & 0 & 0 & 0 & 0 & 0 & 0 & 0 & -1 & 0 & 0 & 0 & 0 & 0 & 0 & 0 \\
 0 & 0 & 0 & 0 & 1 & 0 & 0 & 0 & 0 & 0 & 0 & 0 & 0 & 0 & 0 & 0 \\
 0 & 0 & 0 & 0 & 0 & 0 & 0 & 0 & 0 & 0 & 0 & 0 & 0 & 0 & 0 & 1 \\
 0 & 0 & 0 & 0 & 0 & 0 & 0 & 0 & 0 & 0 & 1 & 0 & 0 & 0 & 0 & 0 \\
 0 & 0 & 0 & 0 & 0 & 1 & 0 & 0 & 0 & 0 & 0 & 0 & 0 & 0 & 0 & 0 \\
 0 & 0 & 0 & 0 & 0 & 0 & 0 & 0 & 0 & -1 & 0 & 0 & 0 & 0 & 0 & 0 \\
 0 & 0 & 0 & 0 & 0 & 0 & 0 & -1 & 0 & 0 & 0 & 0 & 0 & 0 & 0 & 0 \\
 0 & 0 & 1 & 0 & 0 & 0 & 0 & 0 & 0 & 0 & 0 & 0 & 0 & 0 & 0 & 0 \\
 0 & 0 & 0 & 0 & 0 & 0 & 0 & 0 & 0 & 0 & 0 & 0 & 1 & 0 & 0 & 0
 \end{pmatrix}},
\end{equation}
which corresponds to the factorized T-duality along the $T_3$ direction, if the $T_5$ direction is taken to be the M-theory direction.

The dual EDA can be identified as the Type IIB EDA with the structure constants,
\begin{align}
 f'_{12}{}^2& = -1\,,\quad 
 f'_{13}{}^3 = -1\,,\quad
 f'_2{}_{\mathbf{1}}^{34} = 1\,,\quad
 f'_{1\mathbf{1}}{}^{\mathbf{1}} = \tfrac{1}{2}\,,
\nn\\
 Z'_1& = -\tfrac{1}{4}\,,
\end{align}
where for example $f'_2{}_{\mathbf{1}}^{34}$ is a particular component of $f_{\sfa}{}_{\alpha}^{\sfb_1\sfb_2}$\,. 
Parameterizing the group element as $g'= \Exp{x'\,T'_1}\Exp{y'\,T'_2+z'\,T'_3+w'\,T'_4}$\,, the right-invariant 1-form is found as
\begin{equation}
\begin{aligned}
 r' = \rmd x'\,T'_1 + \Exp{-x'}\rmd y\,T'_2 + \Exp{-x'} \rmd z'\, T'_3 + \rmd w' \,T'_4\,.
\end{aligned}
\end{equation}
Signature choice for the generalised metric then must be $\sigma=+1$ and the constant matrix $\hat{\cM}_{AB}\in \text{E}_{5(5)}\times \mathbb{R}^+$. 
Rotating the constant metric by the matrix $C$ defined above one gets the dual constant metric $\hat{\cM}'_{AB}=C_A{}^C\,C_B{}^D\,\hat{\cM}_{CD}$, that is precisely the one obtained by the Buscher's rule, as expected. The dual generalised metric in the Type IIB parametrization gives the following background fields
\begin{equation}
\begin{aligned}
 g_{mn} &= \begin{pmatrix}
 1-4\Exp{-2x}y^2 & 0 & 0 & 2 \Exp{-x} \\
 0 & \Exp{-2x} & 0 & 0 \\
 0 & 0 & 1 & 0 \\
 2 \Exp{-x} & 0 & 0 & 0 
\end{pmatrix} ,\\
 B_2 &= 2\Exp{-2x}y\,\rmd x\wedge\rmd z\,,\qquad
 \Phi = -x\,, 
\end{aligned}
\end{equation}
where $g_{mn}$ is the string-frame metric. 
The external metric in the string frame is simply $g_{\mu\nu}=\diag(1,1,1,1,1,1)$\,. 
This background satisfies the Type IIB supergravity equations of motion.

The transformation discussed above is given by the standard Buscher's rules, but the isometry algebra is non-abelian and we have non-trivial generalised frame fields in the original and the dual frame. In a sense Buscher rules are applied to flat indices rather than to space-time indices. Consequently the map between the two supergravity solutions is indeed non-trivial. 
In particular, in the Type IIB side, the non-standard structure constants such as $f'_{a\alpha}{}^\beta$ and $Z'_a$ are present, this transformation goes beyond the scope of the standard PL T-duality. 

It is worth mentioning, that in the example above the physical algebra in the Type IIB side is non-unimodular. 
In the context of non-abelian T-duality, non-unimodularity of the isometry algebra is known \cite{Fernandez-Melgarejo:2017oyu,Hong:2018tlp,Sakatani:2019jgu,Catal-Ozer:2019hxw,Hlavaty:2019pze} to produce dual backgrounds that respect equations of motion of generalised supergravity \cite{Arutyunov:2015mqj,Wulff:2016tju,Sakamoto:2017wor}. 
However, due to non-trivial $f'_{a\alpha}{}^\beta$ and $Z'_a$ one has $X_{AB}{}^B=0$ and the dual geometry is a solution of the standard supergravity equations of motion.

Note that this background can be regarded as a T-fold if we make an identification $y\sim y+c^y$\,. 
The monodromy matrix is given by a $\beta$-transformation $\Omega_{c^y}=\Exp{c^y R_{\bm{1}}^{34}} \in \text{O}(4,4)\subset \text{E}_{5(5)}$\,. 

\subsection{Generalised Yang--Baxter deformation}

In this section we apply the formalism described above to Yang--Baxter deformations of 10- and 11-dimensional backgrounds. The Yang--Baxter sigma model has been proposed in \cite{Klimcik:2002zj,Klimcik:2008eq} and this has been employed to study various integrable deformations of string theory (see \cite{Orlando:2019his} for a recent review). 
In \cite{Hoare:2016wsk,Borsato:2016pas,Hoare:2016wca}, (a subclass of) Yang--Baxter deformations have been understood as non-abelian T-duals of supergravity backgrounds. 
Indeed, as it has been discussed in \cite{Sakatani:2019zrs}, one can realise Yang--Baxter deformation of group manifold backgrounds as a PL T-duality \eqref{eq:PL-Odd} with the matrix $C_A{}^B$ given by
\begin{align}
 C_A{}^B = \begin{pmatrix} \delta_a^b & 0 \\ r^{ab} & \delta^a_b \end{pmatrix} \in \text{O}(d,d)\,.
\label{eq:YB-C}
\end{align}
For the redefined generators $T'_A$ to form a Drinfel'd double Lie algebra the anti-symmetric constant matrix $r^{ab}$ must satisfy the classical Yang--Baxter equation
\begin{align}
 r^{d a}\,r^{e b}\,f_{de}{}^c + r^{d b}\,r^{e c}\,f_{de}{}^a + r^{d c}\,r^{e a}\,f_{de}{}^b = 0\,.
\end{align}
For this reason such PL T-duality transformation can be called the Yang--Baxter deformation. Under such defined PL T-dualities the generalised metric is transformed as $\cH_{IJ} \to \cH'_{IJ} = (U\,\cH\,U^{t})_{IJ}$ where
\begin{align}
 U_I{}^J(x) \equiv \begin{pmatrix} \delta_m^n& 0 \\ \beta^{mn}(x) & \delta^m_n \end{pmatrix} ,\quad
 \beta^{mn} \equiv r^{ab}\,v_a^m \,v_b^n \,,
\end{align}
and $v_a^m$ are left-invariant vector fields satisfying $[v_a,\,v_b]=f_{ab}{}^c\,v_c$\,. 
Accordingly, this transformation is called the bi-vector deformation (or the $\beta$-deformation) (see for example \cite{Sakamoto:2017cpu,Lust:2018jsx,Sakamoto:2018krs,Catal-Ozer:2019hxw}). 
Since it is a particular class of the PL T-duality, this works as a solution generating transformation in supergravity. Making use of the formulation based on the bi-vector $\b^{mn}$ such deformation can be applied to backgrounds beyond cosets \cite{Bakhmatov:2018apn,Bakhmatov:2018bvp}.

For M-theory, it is natural to expect that this bi-vector deformation can be uplifted to a tri-vector deformation. 
This symmetry was originally explored in \cite{Bakhmatov:2019dow} and earlier in \cite{Lunin:2005jy,Imeroni:2008cr,CatalOzer:2009xd,Deger:2011nb} for abelian U(1)$^3$ deformations. 
The approach to generalised (tri- and hexavector) Yang--Baxter deformations based on the construction of Exceptional Drinfel'd algebras has been developed in \cite{Sakatani:2019zrs,Malek:2019xrf,Malek:2020hpo,Sakatani:2020wah} and a generalisation of the classical Yang--Baxter equation has been proposed. In this formalism Yang--Baxter deformation is defined as a class of E${}_{n(n)}$ rotations which extends the redefinition of generators \eqref{eq:YB-C}.
The associated EDA is called the coboundary EDA, where the dual structure constants, such as $f_a{}^{b_1b_2b_3}$ or $f_a{}^{b_1\cdots b_6}$, are expressed by means of a constant tri-vector $r^{a_1a_2a_3}$ and a hexa-vector $r^{a_1\cdots a_6}$
\begin{equation}
\begin{aligned}
 f_a{}^{b_1b_2b_3} &= 3\,f_{ac}{}^{[b_1|}\,r^{c|b_2b_3]} -3\,Z_a\, r^{b_1b_2b_3}\,,
\\
 f_a{}^{b_1\cdots b_6} &= 6\,f_{ac}{}^{[b_1|}\, r^{c|b_2\cdots b_6]} -10\,f_{a}{}^{[b_1b_2b_3}\, r^{b_4b_5b_6]} \\
 &\quad-6\,Z_a\, r^{b_1\cdots b_6}\,.
\end{aligned}    
\end{equation}
Similar to the case of PL T-duality using the EDA construction one is able to identify the set of the generalised Yang--Baxter equation. 
The explicit form is very complicated for higher exceptional group, but at least for the E${}_{6(6)}$ case, it is proposed in \cite{Malek:2020hpo}. Further in \cite{Gubarev:2020ydf} it has been shown, that beyond coset spaces the same generalised Yang--Baxter equations are the constraint sufficient for tri- and hexavector deformations to generate solutions. 
Earlier in \cite{Bakhmatov:2020kul} some first (to our knowledge) non-trivial examples of non-abelian tri-Killing deformations of 11-dimensional backgrounds, which are not an uplift of 10-dimensional bi-vector deformations have been presented. These were based on the AdS${}_4\times \SS^7$ background. It is important to mention however, that the examples presented there are not generalised Yang--Baxter in the sense, that the corresponding matrix $r^{a_1a_2a_3}$ does not solve the generalised Yang--Baxter equation.

Although tri-vector and hexa-vector deformations can be applied to a general class of backgrounds with at least three Killing vectors, in this paper, we restrict ourselves to the case of group manifolds and provide explicit examples of generalised Yang--Baxter deformations. 

\subsubsection{Tri-vector and hexa-vector in M-theory}

As the first example, let us consider an E$_{6(6)}$ EDA in the M-theory picture generated by the following structure constants
\begin{align}
 f_{24}{}^1 = 1\,,\qquad
 f_{34}{}^2 = 1\,.
\label{eq:alg-YB-M}
\end{align}
Parameterizing the physical group element as $ g = \exp[w T_4]\exp[x T_1 + y T_2 + z T_3 + u T_5 + v T_6]\,,$ the right-invariant 1-form is written as
\begin{align}
 r & = \rmd x\,T_1 + \rmd y\,(T_2 -w\,T_1) + \rmd z\,\bigl(T_3 -w\,T_2 +\tfrac{w^2}{2}\,T_1\bigr) 
\nn\\
 &\quad + \rmd w\,T_4 + \rmd u\,T_5 + \rmd v\,T_6 \,.
\end{align}
The flat metric and the Ricci-flat internal metric are then 
\begin{equation}
\begin{aligned}
 \hat{g}_{ab} &= {\begin{pmatrix}
 0 & 0 & 1 & 0 & 0 & 0 \\
 0 & 1 & 0 & 0 & 0 & 0 \\
 1 & 0 & 0 & 0 & 0 & 0 \\
 0 & 0 & 0 & 1 & 0 & 0 \\
 0 & 0 & 0 & 0 & 1 & 0 \\
 0 & 0 & 0 & 0 & 0 & 1
 \end{pmatrix}},\\
 g_{ij}& = {\begin{pmatrix}
 0 & 0 & 1 & 0 & 0 & 0 \\
 0 & 1 & -2\,w & 0 & 0 & 0 \\
 1 & -2\,w & 2\,w^2 & 0 & 0 & 0 \\
 0 & 0 & 0 & 1 & 0 & 0 \\
 0 & 0 & 0 & 0 & 1 & 0 \\
 0 & 0 & 0 & 0 & 0 & 1\end{pmatrix}} .
\end{aligned}
\end{equation}
We find that the parameterization \eqref{eq:M-param} with $\sigma=+1$ gives the generalised metric $\cM_{IJ}\in \text{E}_{6(6)}\times\mathbb{R}^+$ with $\det\cM_{IJ}=-1$\,. 
The duality-invariant external metric $\mathfrak{g}_{\mu\nu}$ is chosen to proportional to the unity matrix as before. 
Since $\abs{\det (g_{ij})}=1$, the standard external metric $g_{\mu\nu}\equiv \abs{\det (g_{ij})}^{-\frac{1}{3}}\,\mathfrak{g}_{\mu\nu}$ is also flat.
Hence, we start with the original background given by
\begin{align}
 \rmd s^2 = \delta_{\mu\nu}\,\rmd y^\mu\,\rmd y^\nu + g_{ij}\,\rmd x^i\,\rmd x^j\,,
\end{align}
that is a vacuum solution to the Einstein equations. 

To perform redefine of the generators we follow \cite{Malek:2020hpo} (see section 5.4.3), where the coboundary EDA was constructed by twisting the algebra \eqref{eq:alg-YB-M}. Let us consider a particular case where the generalised classical $r$-matrices are given by
\begin{align}
 r^{123} = \frac{1}{\sqrt{2}}\,,\qquad r^{456} = -\sqrt{2}\,,\qquad r^{123456}= \frac{1}{2}\,.
\end{align}
Namely, we consider the following redefinition of generators $T_A\to C_A{}^B\,T_B$ with
\begin{align}
 C = \Exp{\frac{1}{\sqrt{2}}\,R_{123} -\sqrt{2}\,R_{456}}\Exp{\frac{1}{2}\,R_{123456}}\,.
\end{align}
This produces a coboundary EDA with structure constants
\begin{equation}
    \begin{aligned}
 f_{24}{}^1 & = 1\,,&&
 f_{34}{}^2 = 1\,,\\
 f_2{}^{156} &= \sqrt{2}\,,&& f_3{}^{256} = \sqrt{2}\,.
\end{aligned}
\end{equation}
Following the prescription of \cite{Sakatani:2020wah} we can easily find the metric $\cM'_{IJ}$ as
\begin{align}
 \cM'_{IJ} = (U\,\cM\,U^t)_{IJ}\,,
\end{align}
with the local E${}_{n(n)}$ twist matrix $U_I{}^J$ given by
\begin{align}
\begin{split}
 U &= \Exp{\frac{1}{3!}\rho^{i_1i_2i_3}\,R_{i_1i_2i_3}}\Exp{\rho^{123456}\,R_{123456}}
\\
 &=\Exp{\frac{1}{\sqrt{2}} R_{123} - \sqrt{2}\,(R_{456} + y\,R_{156} + z\, R_{256})} \Exp{\frac{1}{2} R^{1\cdots 6}} .
\end{split}
\label{eq:twist-U}
\end{align}
Here, the tri-vector $\rho^{i_1i_2i_3}$ and the hexa-vector $\rho^{123456}$ are defined as \footnote{It is important to mention the relation between our conventions here and those of \cite{Bakhmatov:2020kul,Gubarev:2020ydf}. Tensors $r^{a_1a_2a_3}$ and $r^{a_1\dots a_6}$ here are denoted $\r^{a_1a_2a_3}$ and $\r^{a_1\dots a_6}$ respectively in \cite{Bakhmatov:2020kul,Gubarev:2020ydf}. Tensors $\r^{i_1i_2i_3}$ and $\r^{i_1\dots i_6}$ with space-time indices are denoted $\W^{ijk}$ and $\W^{i_1\dots i_6}$ in \cite{Bakhmatov:2020kul,Gubarev:2020ydf}.}
\begin{equation}
\begin{aligned}
 \rho^{i_1i_2i_3}& \equiv r^{a_1a_2a_3}\,v_{a_1}^{i_1}\,v_{a_2}^{i_2}\,v_{a_3}^{i_3}\,,\\
 \rho^{i_1\cdots i_6} &\equiv r^{a_1\cdots a_6}\,v_{a_1}^{i_1}\cdots v_{a_6}^{i_6}\,,
\end{aligned}    
\end{equation}
where $v_a^i$ are left-invariant vectors
\begin{align}
 v_4 = \partial_w + y\,\partial_x + z \,\partial_y\,,\qquad
 v_a = \partial_a \quad (a\neq 4) 
\end{align}
satisfying $\Lie_{v_a}g_{ij}=0$\,. 
The obtained generalised metric $\cM'_{IJ}(x)$ is such that the signature does not change, i.e. as $\sigma=+1$, and the deformed supergravity fields are
\begin{align}
\begin{split}
 g_{ij} &= -\frac{1}{2(1 + z^2)^{2/3}}{\begin{pmatrix}
 0 & 0 & 4 (1+z^2) & 0 & 0 & 0 \\
   & 4 & -4 (2 w+y z) & -2 z & 0 & 0 \\
   &   & 4 w [4 y z + 2 w (1-z^2)] -4 y^2 & 2 (2 w z-y) & 0 & 0 \\
   &   &   & 1 + 2 z^2 & 0 & 0 \\
   &   &   &   & 1 & 0 \\
   &   &   &   &   & 1 
\end{pmatrix}} , 
\\
 C_3 &= -\sqrt{2}\,\rmd x\wedge\rmd y\wedge\rmd z 
 - \frac{1}{\sqrt{2}\,(1+z^2)}\,\bigl[z\,\rmd y 
 + (y-2 w z)\,\rmd z 
 + \tfrac{\rmd w}{2} \bigr]\wedge \rmd u\wedge\rmd v\,,
\\
 C_6 &= -\frac{1}{4\,(1 + z^2)}\,\rmd x\wedge \cdots \wedge \rmd u\,.
\end{split}
\end{align}
Using the invariance of the external metric $\mathfrak{g}_{\mu\nu}$\,, we find
\begin{align}
 g_{\mu\nu} = (1 + z^2)^{1/3}\,\delta_{\mu\nu}\,.
\end{align}
This is a solution to 11-dimensional supergravity equations of motion. Since the examples of \cite{Bakhmatov:2020kul} do not respect the generalised classical Yang--Baxter equation, the above background gives the first non-trivial example of a non-abelian tri- and hexa-vector Yang--Baxter deformation.

As the twist matrix \eqref{eq:twist-U} indicates, if we make a periodic identification $y\sim y+c^y$\,, this background becomes a U-fold with the monodromy matrix $\Omega_{c^y}=\Exp{-c^y\sqrt{2}\,R_{156}}\in \text{E}_{6(6)}$\,.

\subsubsection{Bi-vector deformation in Type IIB theory}

The tri-vector and hexa-vector deformation tensors $\r^{i_1i_2i_3}$ and $\r^{i_1\dots i_6}$ are dual to the 3-form and 6-form of M-theory in the same sense as the bi-vector $\b^{mn}$ is dual to the Kalb--Ramond field. These correspond to generators of E$_{n(n)}$ of the opposite levels. Since, Type II theory in addition to the NS--NS Kalb--Ramond 2-form contains plenty of RR $p$-form fields originating from the 3-form and 6-form of M-theory, one naturally expects Type II backgrounds to enjoy tri- and hexa-vector deformations as well. Moreover, these must follow the same logic of EDA. Let us consider an explicit example of a generalised Yang--Baxter deformation in Type IIB theory (see \cite{Sakatani:2020wah} for the details).
Start with the following four-dimensional Lie algebra
\begin{align}
 f_{12}{}^3 = 1\,,\qquad
 f_{13}{}^4 = 1\,.
\end{align}
We parameterize the group element as $ g=\Exp{x\,T_1}\Exp{y\,T_2+z\,T_3+w\,T_4}$ and find the right-invariant 1-form to be
\begin{align}
 r_m{}^{\sfa}= {\begin{pmatrix}
  1 & 0 & 0 & 0 \\
  0 & 1 & x & \frac{x^2}{2} \\
  0 & 0 & 1 & x \\
  0 & 0 & 0 & 1 \end{pmatrix}}. 
\end{align}
The constant metric and the Ricci-flat internal metric are then
\begin{equation}
\begin{aligned}
 \hat{g}_{\sfa\sfb}={\begin{pmatrix}
 1 & 0 & 0 & 0 \\
 0 & 1 & 0 & 1 \\
 0 & 0 & 1 & 0 \\
 0 & 1 & 0 & 0 \end{pmatrix}} , &\quad &
 g_{mn}={\begin{pmatrix}
 1 & 0 & 0 & 0 \\
 0 & 2 x^2+1 & 2 x & 1 \\
 0 & 2 x & 1 & 0 \\
 0 & 1 & 0 & 0\end{pmatrix}} .
\end{aligned}
\end{equation}
The external metric $g_{\mu\nu}$ and $\mathfrak{g}_{\mu\nu}$ are chosen as usual as $g_{\mu\nu}=\mathfrak{g}_{\mu\nu}=\diag(1,1,1,1,1,1)$\,. 
The generalised metric $\cM_{IJ}\in \text{E}_{5(5)}\times\mathbb{R}^+$ can be parameterized by using the standard parameterization with $\sigma=+1$\,. 

We find that the E${}_{5(5)}$ transformation $
 C = \Exp{\frac{1}{2!}\,r_{\alpha}^{\sfa\sfb}\,R_{\sfa\sfb}^{\alpha}},
$
with
\begin{equation}
\begin{aligned}
 r_{\mathbf{1}}^{23} &= p_1\,,&&
 r_{\mathbf{1}}^{14} = p_1\,q_1\,,\\
 r_{\mathbf{1}}^{24} &= p_1\,q_2\,,&&
 r_{\mathbf{2}}^{23} = p_2\,,\\
 r_{\mathbf{2}}^{14} &= p_2\,q_1\,,&&
 r_{\mathbf{2}}^{24} = p_2\,q_2\,,
\end{aligned}    
\end{equation}
maps the original EDA to a coboundary EDA with
\begin{equation}
\begin{aligned}
 f_{12}{}^3 &= 1\,,\quad
 f_{13}{}^4 = 1\,,\\
 f^{\mathbf{1}}_1{}^{24}& = p_1\,,\quad
 f^{\mathbf{1}}_1{}^{34} = p_1\,q_2\,,\quad
 f^{\mathbf{1}}_2{}^{34} = - p_1\,q_1 \,, 
\\
 f^{\mathbf{2}}_1{}^{24} &= p_2\,,\quad
 f^{\mathbf{2}}_1{}^{34} = p_2\,q_2\,,\quad
 f^{\mathbf{2}}_2{}^{34} = -p_2\,q_1\,.
\end{aligned}    
\end{equation}
The components $r_{\alpha}^{\sfa\sfb}$ can then be understood as a generalisation of the classical $r$-matrix components. 

Let us choose for simplicity $p_1=p_2=\eta$, $q_1=-1$, and $q_2=0$ for which 
\begin{equation}
   \begin{aligned}
 C& = \Exp{\eta\,(R^{\mathbf{1}}_{23} + R^{\mathbf{2}}_{23} - R^{\mathbf{1}}_{14} - R^{\mathbf{2}}_{14})}\,, \\
 f_{12}{}^3 &= f_{13}{}^4 = 1\,,\\
 f^{\mathbf{1}}_1{}^{24}& = f^{\mathbf{1}}_2{}^{34} = f^{\mathbf{2}}_1{}^{24} =f^{\mathbf{2}}_2{}^{34} = \eta \,.
\end{aligned} 
\end{equation}
In this case, the generalised Yang--Baxter deformation map reads
\begin{equation}
\begin{aligned}
 \cM_{IJ} &\to (U\,\cM\,U^t)_{IJ}\,,\\
 U &=\Exp{\eta\,[R^{\mathbf{1}}_{23} + R^{\mathbf{2}}_{23} - (R^{\mathbf{1}}_{14} + R^{\mathbf{2}}_{14} - y\, (R^{\mathbf{1}}_{34} + R^{\mathbf{2}}_{34})]}\,.
\end{aligned}    
\end{equation}

From such obtained deformed metric in the Type IIB parametrisation one reads the following 10-dimensional Einstein-frame metric and gauge fields
\begin{align}
 \mathsf{g}_{mn} &= 
 \frac{2\,\eta^2}{[1+2\,\eta^2\,(1-2x^2-2y)-4\,\eta^4]^{\frac{3}{4}}} \times 
\nn\\
 \multispan2{
 $\times\begin{pmatrix}
 \frac{1}{2\,\eta^2}-2 x^2-2 y+1 & -2 x & -1 & 0 \\
 -2 x & \frac{x^2+\frac{1}{2}}{\eta^2}-y^2-1 & \frac{x}{\eta^2} & \frac{1}{2\,\eta^2}-y \\
 -1 & \frac{x}{\eta^2} & \frac{1}{2\,\eta^2} & 0 \\
 0 & \frac{1}{2\,\eta^2}-y & 0 & -1 \end{pmatrix}$
 },
\nn\\
 \mathsf{g}_{\mu\nu} &= \bigl[1 + 2\,\eta^2\,(1 - 2x^2 - 2y) -4\,\eta^4\bigr]^{\frac{1}{4}}\times 
\nn\\
  &\quad \diag(1,1,1,1,1,1)\,,
\nn\\
 \Exp{-2\Phi} &= \dfrac{1 + 2\eta^2\,(1 - 2x^2 - 2y) - 4 \eta^4}{\bigl[1 + \eta^2\,(1 - 2x^2 - 2y) - 2\eta^4\bigr]^2} \,,
\\
 C_0  &= \frac{1}{1-2 \eta ^4+\eta^2\,(1-2x^2-2y)}-1 \,,
\nn\\
 C_2 &= \eta(1 + 2\,\eta^2\,(1-2x^2 - 2y) - 4\eta^4)^{-1}\times 
\nn\\
  &\quad {(1 - 2\,\eta^2\,y)\, \rmd x\wedge \rmd y - (1 - 2x^2 - y - 2\eta^2)\,\rmd y\wedge \rmd z}
\nn\\
 &\quad +\frac{\eta\,\bigl(- 2\eta^2\,\rmd x + 2\,x\,\rmd y + \rmd z \bigr)\wedge \rmd w}{1 + 2\,\eta^2\,(1-2x^2 - 2y) - 4\eta^4} = -B_2 \,.\nn
\end{align}

One finds that the deformation is well defined for small enough values of the deformation parameter $\eta$ (e.g., $\eta=1/2$ works). In this region the metric has real value in the vicinity of the origin and one checks that this is indeed a solution of Type IIB supergravity equations of motion.

If we make a periodic identification $y\sim y+c^y$\,, this background becomes a U-fold with the monodromy matrix $\Omega_{c^y}=\Exp{c^y \eta\,(R_{\bm{1}}^{34}+R_{\bm{2}}^{34})}\in \text{E}_{5(5)}$\,. 
This is a mixture of the $\beta$-transformation and the $\gamma$-transformation and is not a T-fold. 

\section{Conclusions}
\label{sec:concl}

In this paper we constructed various 11D/10D solutions using the procedure of non-abelian U-duality. The formalism has been developed in great details in a number of works \cite{Sakatani:2019zrs,Malek:2019xrf,Sakatani:2020wah,Malek:2020hpo,Blair:2020ndg,Sakatani:2020iad} and so far has not been enriched by a single non-trivial example. In this work we consider a set of backgrounds of 11-dimensional supergravity, synthetic in the sense that these are Ricci flat, and some are just a Minkowski space. Most of the examples of non-abelian U-duality transformations we find are beyond the realm of the PL T-duality. We consider both transformations mapping backgrounds of M-theory and those that relate M-theory and Type IIB backgrounds. For the former we were forced to choose a specific signature for the generalised metric of the corresponding exceptional field theory such that the obtained backgrounds are solutions of M$^*$- or Type II$^*$ theories. Hence, the presented examples of non-abelian U-dualities contain T-duality reflection of time, which we show explicitly for the duality relating M-theory and Type IIB* theory backgrounds. Also 
we study the cases where the EDA's are of the coboundary type and the corresponding non-abelian duality gives generalised Yang--Baxter deformations. 
Examples of generalised Yang--Baxter deformations both in M-theory and Type IIB theory are presented. To our knowledge, these are the first examples of tri- and hexa-vector generalised Yang--Baxter deformations presented in the literature. It is worth mentioning here the results of \cite{Bakhmatov:2020kul}, where non-abelian tri-Killing deformations of AdS$_4\times \SS^7$ within properly truncated SL(5) exceptional field theory have been found. However, as it has been noticed in \cite{Gubarev:2020ydf} the corresponding $r$-tensor (in our notations here) does not satisfy the generalised Yang--Baxter equation, that is for the SL(5) case simply the unimodularity constraint. Hence, these deformations are tri-Killing but not generalised Yang--Baxter, which illustrates that this algebraic condition is only sufficient for a deformation to give a solution.

The restriction of our results is that considered were only EDA's with unimodular structure constants $X_{AB}{}^B=0$ (i.e., EDA's with vanishing trombone gauging). 
For a non-unimodular EDA one gets a background which does not satisfy the supergravity equations of motion. 
Indeed, we were not able to find any supergravity solutions for non-unimodular EDA. 
As it has been discussed in the context of PL T-duality \cite{Demulder:2018lmj,Sakatani:2019jgu}, for the non-unimodular, the obtained background will be a solution of certain deformed supergravity. 
It is interesting to identify what kind of deformed supergravity is needed to in order to make the dual background to be a solution.
In order to clarify this point, it will be useful to study the flux formulation of exceptional field theory and express the equations of motion in terms of generalised fluxes.

Another restriction we have been put on comes from the construction of exceptional Drinfel'd algebra, that is the original background must be a group manifold.
Given this issue, we were not able to study non-abelian U-duality transformations of more interesting backgrounds, such as with AdS factors, relevant to holography in M-theory. For that one has to formulate the procedure of non-abelian U-duality for coset spaces, that for non-abelian T-duality has been done in \cite{Alvarez:1994np,Lozano:2011kb} and for PL T-duality in \cite{Demulder:2019vvh}. It would be important to study an extension of their analysis to the case of non-abelian U-duality.

\subsection*{Acknowledgments}

The work of ETM was supported by the Foundation for the Advancement of Theoretical Physics and Mathematics ``BASIS'', by Russian Ministry of education and science and in part by the program of competitive growth of Kazan Federal University.
The work of YS was supported by JSPS Grant-in-Aids for Scientific Research (C) 18K13540 and (B) 18H01214.

\appendix

\section{Exceptional Drinfeld algebras}\label{app:edas}

Given generators of the isotropic subalgebra $\{T_a\}=\bas \gg$ with $a=1,\dots,n$, one denotes additional generators of the exceptional Drinfeld algebra as $\{T^{a_1a_2}, T^{a_1,\dots a_5}\}$. Its algebraic structure is then defined by specifying multiplication rules, which for $n\leq 6$ following  \cite{Malek:2020hpo} can be written as:
\begin{equation}
\label{eq:eda6}
\begin{aligned}
 T_a \circ T_b =&\ f_{ab}{}^c\,T_c \,,
\\
 T_a \circ T^{b_1b_2} &= f_a{}^{b_1b_2c}\,T_c + 2\,f_{ac}{}^{[b_1}\,T^{b_2]c}
 +3\,Z_a\,T^{b_1b_2}\,,
\\
 T_a \circ T^{b_1\cdots b_5} &= -f_a{}^{b_1\cdots b_5c}\,T_c - 10\,f_{a}{}^{[b_1b_2b_3}\,T^{b_4b_5]} - 5\,f_{ac}{}^{[b_1}\,T^{b_2\cdots b_5]c} 
 +6\,Z_a\,T^{b_1\cdots b_5}\,,
\\
T^{a_1a_2} \circ T_b &= -f_b{}^{a_1a_2c}\,T_c + 3\,f_{[c_1c_2}{}^{[a_1}\,\delta^{a_2]}_{b]}\,T^{c_1c_2} -9\,Z_c\,\delta_b^{[c}\,T^{a_1a_2]}\,,\\
 T^{a_1a_2} \circ T^{b_1b_2} &= -2\, f_c{}^{a_1a_2[b_1}\, T^{b_2]c} - f_{c_1c_2}{}^{[a_1}\,T^{a_2]b_1b_2c_1c_2} +3\,Z_c\,T^{a_1a_2b_1b_2c}\,,
\\
 T^{a_1a_2} \circ T^{b_1\cdots b_5} &= 5\,f_c{}^{a_1a_2[b_1}\, T^{b_2\cdots b_5]c} \,,
\\
 T^{a_1\cdots a_5} \circ T_b &= f_b{}^{a_1\cdots a_5c}\,T_c + 10\,f_b{}^{[a_1a_2a_3}\,T^{a_4a_5]}  + 20\,f_c{}^{[a_1a_2a_3}\,\delta_b^{a_4}\,T^{a_5]c} 
 + 5\,f_{bc}{}^{[a_1}\,T^{a_2\cdots a_5]c} \\
 &\quad+ 10\,f_{c_1c_2}{}^{[a_1}\,\delta^{a_2}_b\,T^{a_3a_4a_5]c_1c_2} 
 -36\,Z_c\,\delta_b^{[c}\,T^{a_1\cdots a_5]}\,,
\\
 T^{a_1\cdots a_5} \circ T^{b_1b_2} &= 2\,f_c{}^{a_1\cdots a_5[b_1}\,T^{b_2]c} - 10\,f_c{}^{[a_1a_2a_3}\, T^{a_4a_5]b_1b_2c}\,,
\\
 T^{a_1\cdots a_5} \circ T^{b_1\cdots b_5} &= -5\,f_c{}^{a_1\cdots a_5[b_1}\, T^{b_2\cdots b_5]c} \,.
\end{aligned}
\end{equation}

Choosing an $(n-1)$-dimensional maximally isotropic algebra, the EDA is naturally expressed in the Type IIB language. For $n\leq 5$ the Type IIB EDA becomes\cite{Sakatani:2020wah}
\begin{align}
\label{eq:edaB}
\begin{split}
 T_{\sfa}\circ T_{\sfb} &=f_{\sfa\sfb}{}^{\sfc}\,T_{\sfc}\,,
\\
 T_{\sfa}\circ T^{\sfb}_\beta &= f_{\sfa}{}_{\beta}^{\sfc\sfb}\,T_{\sfc}
 + f_{\sfa\beta}{}^{\gamma}\,T_\gamma^{\sfb} - f_{\sfa\sfc}{}^{\sfb}\,T_\beta^{\sfc} 
 +2\,Z_{\sfa}\,T^{\sfb}_\beta \,,
\\
 T_{\sfa}\circ T^{\sfb_1\sfb_2\sfb_3} &=
 f_{\sfa}{}^{\sfc\sfb_1\sfb_2\sfb_3}\, T_{\sfc} + 3\,\epsilon^{\gamma\delta}\,f_{\sfa}{}_{\gamma}^{[\sfb_1\sfb_2}\, T_{\delta}^{\sfb_3]} - 3\,f_{\sfa\sfc}{}^{[\sfb_1}\, T^{\sfb_2\sfb_3]\sfc} 
 +4\,Z_{\sfa}\,T^{\sfb_1\sfb_2\sfb_3}\,,
\\
 T^{\sfa}_\alpha\circ T_{\sfb} &= 
 f_{\sfb}{}_{\alpha}^{\sfa\sfc} \, T_{\sfc} 
 + 2\,\delta^{\sfa}_{[\sfb}\,f_{\sfc]\alpha}{}^{\gamma}\, T_\gamma^{\sfc}
 + f_{\sfb\sfc}{}^{\sfa}\,T_\alpha^{\sfc} 
 +4\,Z_{\sfc}\,\delta^{[\sfa}_{\sfb}\,T^{\sfc]}_\alpha\,,
\\
 T^{\sfa}_\alpha\circ T^{\sfb}_\beta &= - f_{\sfc}{}_{\alpha}^{\sfa\sfb}\,T_\beta^{\sfc} 
 - f_{\sfc\alpha}{}^\gamma\,\epsilon_{\gamma\beta}\,T^{\sfc \sfa\sfb} 
 + \tfrac{1}{2}\,\epsilon_{\alpha\beta}\,f_{\sfc_1\sfc_2}{}^{\sfa}\,T^{\sfc_1\sfc_2\sfb}
 -2\,\epsilon_{\alpha\beta}\,Z_{\sfc}\,T^{\sfa\sfb\sfc} \,,
\\
 T^{\sfa}_\alpha\circ T^{\sfb_1\sfb_2\sfb_3}
 &= -3\,f_{\sfc}{}_{\alpha}^{\sfa[\sfb_1}\,T^{\sfb_2\sfb_3]\sfc} \,,
\\
 T^{\sfa_1\sfa_2\sfa_3} \circ T_{\sfb} 
 &= -f_{\sfb}{}^{\sfc \sfa_1\sfa_2\sfa_3}\,T_{\sfc} 
 - 6\,\epsilon^{\gamma\delta}\,f_{[\sfb|}{}_{\gamma}^{[\sfa_1\sfa_2}\,\delta_{|\sfc]}^{\sfa_3]}\,T_\delta^{\sfc}
\\
 &\quad + 3\,f_{\sfb\sfc}{}^{[\sfa_1}\, T^{\sfa_2\sfa_3]\sfc}
 + 3\, f_{\sfc_1\sfc_2}{}^{[\sfa_1}\,\delta_{\sfb}^{\sfa_2}\,T^{\sfa_3]\sfc_1\sfc_2}
 +16\,Z_{\sfc}\,\delta_{\sfb}^{[\sfa_1}\,T^{\sfa_2\sfa_3\sfc]} \,,
\\
 T^{\sfa_1\sfa_2\sfa_3} \circ T_\beta^{\sfb} 
 &= -f_{\sfc}{}^{\sfa_1\sfa_2\sfa_3 \sfb}\,T_\beta^{\sfc}
  + 3\, f_{\sfc}{}_{\beta}^{[\sfa_1\sfa_2}\,T^{\sfa_3] \sfb\sfc} \,,
\\
 T^{\sfa_1\sfa_2\sfa_3} \circ T^{\sfb_1\sfb_2\sfb_3} 
 &= -3\, f_{\sfc}{}^{\sfa_1\sfa_2\sfa_3 [\sfb_1}\, T^{\sfb_2\sfb_3] \sfc} \,.
\end{split}
\end{align}
The indices $\sfa,\sfb$ run over $\sfa,\sfb=1,\dotsc,n-1$ and $\alpha,\beta =\mathbf{1},\mathbf{2}$ label doublets of the $\text{SL}(2)$ $S$-duality. 
When $n=4$, the number of the generators $T_A=\bigl(T_{\sfa},\,T_\alpha^{\sfa},\,\frac{T^{\sfa_1\sfa_2\sfa_3}}{\sqrt{3!}}\bigr)$ is 10 and matches with the M-theory EDA.

\bibliographystyle{utphys}
\bibliography{biblio.bib}

\end{document}